%% file: main-qce-2026.tex
\documentclass[conference]{IEEEtran}
\IEEEoverridecommandlockouts
\usepackage{cite}
\usepackage{amsmath,amssymb,amsfonts}
\usepackage{graphicx}
\usepackage{textcomp}
\usepackage{xcolor}
\usepackage{comment}
\usepackage[hyphens]{url}

\usepackage[capitalise, nameinlink]{cleveref}
\usepackage[T1]{fontenc}

\usepackage[inline]{enumitem}
\usepackage{physics2}
\usephysicsmodule{ab,ab.braket}
\usepackage{braket}
\usepackage{booktabs}

\PassOptionsToPackage{compatibility=false}{caption}
\usepackage{subcaption}

\usepackage{blindtext}

\usepackage{algpseudocodex}
\usepackage{algorithm}

\usepackage{mathtools}

\usepackage{multicol}

\DeclareMathOperator*{\argmin}{arg\,min}

\def\BibTeX{{\rm B\kern-.05em{\sc i\kern-.025em b}\kern-.08em
    T\kern-.1667em\lower.7ex\hbox{E}\kern-.125emX}}
\begin{document}

\title{SEQC: Stratify-Elaborate Quantum Compilation Towards Modular Hybrid Architectures
\thanks{This work was partially supported by the U.S. Department of Energy, Office of Science, National Quantum Information Science Research Centers, Superconducting Quantum Materials and Systems Center (SQMS), under Contract No. 89243024CSC000002. Fermilab is managed by FermiForward Discovery Group, LLC, acting under Contract No. 89243024CSC000002.

This research was partially supported by NSF awards CNS-1763743
and CCF-2119069, Academic Year Undergraduate Research Award
\#2023-17704 by Northwestern University, and undergraduate summer research awards by the Robert R. McCormick School of Engineering and Applied Science and the Computer Science Department at Northwestern University. 

This research was supported in part through the computational
resources and staff contributions provided for the Quest high performance computing facility at Northwestern University which
is jointly supported by the Office of the Provost, the Office for
Research, and Northwestern University Information Technology.
We acknowledge the use of IBM Quantum services for this work.
The views expressed are those of the authors, and do not reflect the
official policy or position of IBM or the IBM Quantum team.
}
}

\author{\IEEEauthorblockN{Mingyoung Jessica Jeng$^{1\dagger*}$, Nikola Vuk Maruszewski$^{2\dagger*}$, Mu-Te Lau$^{1*}$, Connor Selna$^{1*}$\\ Michael Gavrincea$^3$,
Kaitlin N. Smith$^{1**}$, Nikos Hardavellas$^{1**}$}
\IEEEauthorblockA{\textit{$^1$Department of Computer Science, Northwestern University, Evanston, IL, USA} \\
\textit{$^2$College of Computing, Georgia Institute of Technology, Atlanta, GA, USA}\\
\textit{$^3$MIT Lincoln Laboratory, Lexington, MA, USA}\\
$^{\dagger}$These authors contributed equally.
$^{*}$mingyoungjeng@u.northwestern.edu, $^{2}$nikola@gatech.edu, \\
$^{*}$\{mtlau, ConnorSelna2021\}@u.northwestern.edu; 
$^{3}$mgav@mit.edu,
$^{**}$\{kns, nikos\}@northwestern.edu
}
}

\definecolor{nudarkred}{HTML}{D85820}
\definecolor{nubrightred}{HTML}{EF553F}

\bstctlcite{IEEEexample:BSTcontrol}

\maketitle

\begin{abstract}
As quantum computing technology matures, the pursuit of performance and scalability has led to the widespread adoption of modular quantum architectures. We expect that the next stage of technological evolution will integrate multiple qubit modalities into these systems, producing hybrid, modular quantum architectures. However, the complexity of hybrid, modular quantum devices, coupled with their growing sizes, presents an imminent scalability challenge for quantum compilation. Existing qubit allocation methods are often unable to contend with inter-module links, which do not necessarily support a universal basis gate set. Furthermore, these algorithms are typically not designed for qubit links of significantly varying latency or fidelity. In this work, we propose SEQC, a hierarchical parallelized compilation pipeline optimized for modular quantum systems, including several novel methods for qubit placement, qubit routing, and circuit optimization. SEQC attains a 9.3--32.3\% average increase in circuit fidelity (49.99--63.36\% max), depending on the chiplet size and topology. Additionally, owing to its ability to parallelize compilation, SEQC achieves 1.34--3.27$\times$ faster compilation on average (3.37--6.74$\times$ max) over a chiplet-unaware Qiskit baseline.
\end{abstract}


\input{2026_04_24_qce}

\section*{Acknowledgment}
The authors thank Kabir Dubey for helping with earlier versions of this manuscript.
\clearpage

\bibliographystyle{IEEEtran}
\bibliography{refs}

\end{document}

%% file: 2026_04_24_qce.tex
\section{Introduction} \label{sec:intro}

In classical computing, physical, technological, and economic constraints have prevented single monolithic systems from scaling to the point that they can meet the demand for high performance.
In response, classical computer systems have become increasingly distributed. 
We postulate that quantum computing is on a similar path. 
Practical implementations of quantum computation will require millions of physical qubits~\cite{paler2019really}, much more than are likely to fit in a single die~\cite{vanMeter_2010_DQC,smith_2022_chiplets}.
While a distributed network of quantum computers is still far away, we are already observing the first steps to displace monolithic QPUs and adopt a modular architecture.
Recent developments in quantum chip linking, including flip-chip architectures~\cite{gold2021entanglement} and low-loss coaxial cables~\cite{niu2023low}, suggest that modular designs are the most viable for scaling~\cite{smith_2022_chiplets}.
Many contemporary or upcoming leading quantum systems adopt chiplet-based modular QPUs supported by a variety of links, including
high-quality but short-range ones~\cite{Rigetti_AspenM, gold2021entanglement,gambetta_2024_IBMcouplers, Mandelbaum_2024_IBMCouplers}; lower-quality but longer-range~\cite{gambetta_2024_IBMcouplers, Mandelbaum_2024_IBMCouplers}; a combination of both~\cite{Mandelbaum_2024_IBMCouplers}; or multi-chip connectivity through tunable couplers and routing chips~\cite{Rigetti_Cepheus, Field_2024_RigettiModular}. 
Modular QPUs dominate the latest roadmaps of several major companies~\cite{IBM_2025_FTQC, IonQ_2024_Roadmap, Rigetti_2024_Roadmap}, as they are expected to allow them to meet quantum scaling targets in practical ways.

At the same time, no single qubit technology simultaneously offers long coherence, fast gates, high connectivity, and scalable, mature fabrication. 
Superconducting qubits, for example, excel at gate latencies and  fabrication maturity, but suffer from limited connectivity and short coherence times. 
Trapped ions have excellent coherence, but slow speeds, and all-to-all connectivity only within a single trap segment. 
Hybrid modular quantum architectures provide a path to combine these technologies and enable performance impossible for any single modality alone.
As such,
they are increasingly recognized as a pragmatic design path for long-term universal quantum computers~\cite{DARPA_HARQ, HQAN, hetarch, Xiang_superconducting_hybrid_2013}.

Unfortunately, current compilation infrastructure is incapable of addressing heterogeneous quantum systems. Efforts to realize these systems
have predominantly leveraged quantum state transfer or SWAPs between heterogeneous qubit technologies~\cite{Kurizki_hybrid_2015, Clerk2020, Zhou2014, PhysRevLett.108.130504}.
While entangling quantum operations, e.g., CNOT or CZ, 
between different
technologies have been demonstrated~\cite{PhysRevLett.114.080501, Xiang_superconducting_hybrid_2013, Cia_2024, Daniilidis_2013, Yu2016}, 
they often compromise on gate fidelity, and have yet to be realized between some common qubit technologies, e.g., superconducting and trapped-ion qubits. 
Thus, inter-module links in hybrid modular systems may not be able to support universal gate sets.
Moreover, the run time of state-of-the-art, monolithic compilation methods scales quadratically with the number of qubits $n$ in the quantum processor (experimentally observed in \cref{fig:10Q_stratification_time}). 
While theoretically polynomial, an $O(n^2)$ scaling becomes prohibitively expensive as the number of qubits grows.
To make matters more challenging, qubits, couplers, and gates have diverse error profiles that are highly variable both spatially and temporally~\cite{Tannu_2019_ErrorVariability, Murali_2019_NoiseMapping, dasgupta2021stability}. 
Owing to this, unlike classical compilation that is done only once for an architecture, quantum programs must be recompiled every time before execution, especially across calibrations windows.

To address these issues, in this paper we leverage hardware modularity to achieve modular compilation.
Inspired by compilation in classical systems, which typically runs independently and in parallel for each source file,
we propose a compilation framework for modular quantum processors that \textbf{stratifies} the source quantum circuit, i.e., splits it into subcircuits that fit in each module, maps subcircuits to modules, 
and then \textit{in parallel} \textbf{elaborates} each subcircuit to compile it to its target module. This \textbf{Stratify-Elaborate Quantum Compiler (SEQC)} stratifies a source program only once for a given modular architecture, and performs only the elaboration step recurrently before each execution. 
This replaces a global $O(n^2)$ recompilation with several parallelizable $O(k^2)$ elaboration steps for $k$-qubit modules. As $k$ is expected to remain relatively stable or grow at a much slower pace in the foreseeable future~\cite{IBM_2025_FTQC}, 
SEQC's recompilation latency will barely grow as processors scale.

Additionally, SEQC tailors to hardware modularity by minimizing inter-module communication. 
Compared to today's stock compilers, SEQC yields circuits with shorter execution times, significantly fewer inter-module gates, and ultimately higher execution fidelity. 
More importantly, as the number of qubits in a processor grows, SEQC achieves even higher performance in these figures of merit.

In summary, the contributions of this paper are as follows:
\begin{itemize}
    \item We make stock compilers aware of modularity 
    through a peephole correction pass,
    thereby allowing them to correctly compile circuits for modular architectures with limited inter-module gate support.
    \item We design and implement SEQC, a Stratify-Elaborate Quantum Compiler for modular architectures. By limiting recompilation between runs to within individual modules, SEQC's compilation time is largely unaffected by the growth of qubit counts in future quantum processors.
    \item We design and implement in SEQC several novel methods for qubit placement, qubit routing, and circuit optimization tailored for hybrid modular compilation, including a SWAP classification framework that facilitates reducing inter-module gates, a qubit-pinning technique that ensures subcircuit modularity, and stratification based on simulated annealing and spatio-temporal aware (STA) qubit allocation techniques.
    \item We evaluate SEQC on a modular quantum system consisting of small-scale (10-qubit heavy-hex) modules, and find that it compiles circuits with $9.3\%$ higher circuit fidelity on average (up to $49.99\%$), while consistently achieving faster compilation times ($3.27\times$ on average, up to $6.74\times$) compared to a module-unaware Qiskit baseline. 
    When evaluated on a system with large-scale (120-qubit grid topology) modules, SEQC achieves $32.3\%$ higher fidelity on average (up to $63.36\%$), while compiling $1.34\times$ faster on average (up to $3.37\times$). More importantly, SEQC outperforms Qiskit even when the latter is afforded significant allowances (no universal gate restrictions, and much lower inter-module gate errors).
\end{itemize}

\section{Background and Motivation} \label{sec:background}

Compilation is often divided into distinct qubit allocation (layout and routing), basis translation, 
and optimization stages.

\paragraph{Qubit Allocation} \label{sec:background_qubit_allocation}

Qubit allocation addresses the topology constraints of quantum devices, where physical qubits on a device have limited connectivity to other qubits on the device. 
It is critical that after this stage, which often modifies or adds circuit gates,
the resulting, technology-dependent circuit is functionally equivalent to the original, technology-independent algorithm~\cite{smith2019quantum}. 

Qubit allocation is NP-complete~\cite{qubit_allocation}, and typically divided into two substages to make it more tractable: \emph{qubit layout} assigns initial positions of virtual qubits on the device ~\cite{qubit_routing, sabre_swap}, while \emph{qubit routing} operates from this initial layout and inserts SWAP operations to align a quantum circuit's gates to a given qubit topology.





Qubit routing, which remains NP-hard~\cite{qubit_routing}, is often simplified by assuming qubit links are equivalent, particularly with respect to average fidelity and two-qubit gate duration. 
%
However, for modular architectures, the inter-module connections are expected to be far worse compared to intra-module connections~\cite{smith_2022_chiplets, PhysRevLett.114.080501,
Zhou2014,
PhysRevLett.108.130504,
Yu2016}.
Thus, as we will show, the current simplified model applied on modular architectures leads to worse solutions than a more physically accurate one.

\paragraph{Basis Translation}

Basis translation converts the gates of an input quantum circuit into gates that are physically supported on hardware. 
Basis translation is a well-studied and efficiently solved compiler task given a universal target gate set~\cite{mike_and_ike,Solovay-Kitaev,qiskit2024}.
Unfortunately, modular quantum architectures depend upon inter-module links that may only support data movement (e.g., SWAPs) and not universal operations~\cite{niu2023low}. 
Thus, compilers that depend on the expectation of universality among qubit links~\cite{qiskit_basis_translation_errors}, and do not expect these \emph{heterogeneous, non-universal} basis gate sets, can generate incompatible outputs when targeting modular architectures.

\paragraph{Optimization}

Quantum circuit optimization seeks to prune extraneous operations by combining, eliminating, and parallelizing quantum gates~\cite{optimization1}. 
Optimization is highly time and resource intensive, with the problem scaling exponentially to the number of qubits and circuit depth~\cite{qfactor}. As a result, partitioning, the division of a quantum circuit into \emph{blocks} of quantum gates, 
is an optimization technique~\cite{qfactor, qgo, bqskit}. 
One major consideration in this process is where and when to draw the line between blocks, as optimization opportunities are forfeited at these boundaries, which must be carefully chosen to minimize the negative impact on the output solution. 
Modular architectures are naturally synergistic with partitioning, since inter-module connections present physically motivated boundaries between operations.

\section{Compilation for Hybrid Modular Quantum Systems} \label{sec:design}

While SEQC supports hybrid modular systems with heterogeneous QPUs, this paper focuses on the homogeneous case. This allows us to isolate the impact of two-stage compilation from the effects of QPU heterogeneity. It also enables fair comparison with existing compilers, none of which, to our knowledge, target multiple qubit modalities simultaneously. Extending SEQC to hybrid systems requires only modality-specific elaboration stages; the stratification stage and other core techniques remain largely unaffected.
As is often the case in hybrid quantum systems, our model of modular hybrid quantum architectures considers \emph{non-universal} inter-module links. In particular, we assume they only support SWAPs~\cite{niu2023low}.
Many compilers (including those for modular architectures, see \cref{sec:related_work}) cannot handle non-universal inter-module links. For example, the Qiskit compiler~\cite{qiskit2024} only supports universal basis gate sets~\cite{qiskit_basis_translation_errors}; its routing stage~\cite{zou2024lightsabrelightweightenhancedsabre} (an implementation of SABRE) will place gates on incompatible links with no mechanism to detect or correct the resulting errors. Moreover, as devices scale, compilers must balance the contending figures of solution quality and compilation latency.
To ameliorate these issues, we present
two module-aware compiler frameworks: a peephole compilation pass (\cref{sec:peephole}) that adds support for non-universal module links to \emph{any} compiler, and the stratify-elaborate quantum compiler (SEQC, \cref{sec:seqc}), which advocates for a hierarchical two-stage compilation that produces better circuits, faster.

\subsection{Module-awareness via Peephole Correction} \label{sec:peephole}

\begin{figure*}[tbp]
    \centering
    \begin{subfigure}[t]{0.22\linewidth}
        \centering
        \includegraphics[height=7em]{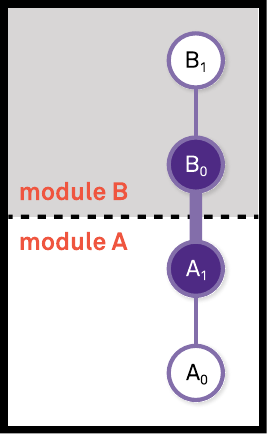}
        \captionsetup{justification=centering, singlelinecheck=false}
        \caption{Example 2-module\\architecture}
        \label{fig:peephole_topology}
    \end{subfigure}
    \hfill
    \begin{subfigure}[t]{0.22\linewidth}
        \centering
        \includegraphics[height=7em]{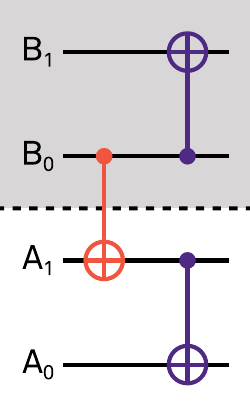}
        \captionsetup{justification=centering, singlelinecheck=false}
        \caption{Input circuit from\\qubit allocation}
        \label{fig:peephole_initial}
    \end{subfigure}
    \hfill
    \begin{subfigure}[t]{0.26\linewidth}
        \centering
        \includegraphics[height=7em]{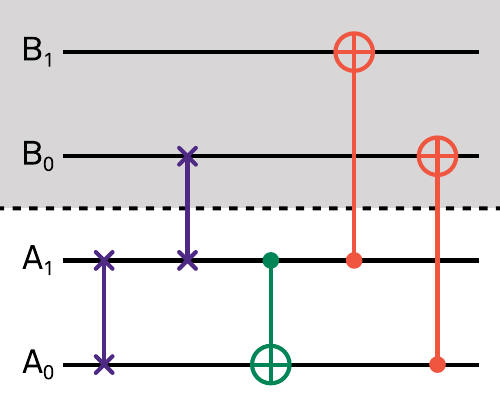}
        \captionsetup{justification=centering, singlelinecheck=false}
        \caption{Insertion of SWAPs\\for cyclic qubit rotation}
        \label{fig:peephole_swap}
    \end{subfigure}
    \hfill
    \begin{subfigure}[t]{0.28\linewidth}
        \centering
        \includegraphics[height=7em]{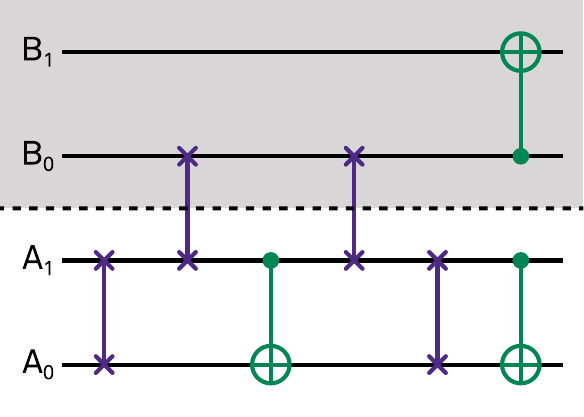}
        \captionsetup{justification=centering, singlelinecheck=false}
        \caption{Insertion of return SWAPs}
        \label{fig:peephole_undo}
    \end{subfigure}
    \caption{%
    Peephole correction for inter-module gates that are unimplementable on hardware (i.e., non-SWAP, shown in \textcolor{nubrightred}{orange}).
    }
    \label{fig:peephole}
\end{figure*}

Our peephole correction pass (\cref{fig:peephole}) detects non-SWAP gates placed on non-universal inter-module links in a post-qubit allocation circuit (\cref{fig:peephole_initial}), and uses SWAPs to move them to nearby intra-module links through a cyclic qubit rotation (\cref{fig:peephole_swap}). 
Additional ``return SWAPs'' are then inserted to return the qubits to their initial locations (\cref{fig:peephole_undo}).
Overall, for every inter-module gate in the original circuit generated by SABRE, the peephole pass inserts 4 additional SWAP gates---2 inter-module and 2 intra-module. 
We note that the overhead of return SWAPs is inevitable when the stock compiler is unaware of module-specific device constraints, as the disruption in qubit layout can quickly cascade and deviate from the given routing.


\subsection{The Stratify-Elaborate Quantum Compiler} \label{sec:seqc}

\begin{figure*}[ht]
    \centering
    \includegraphics[width=\textwidth]{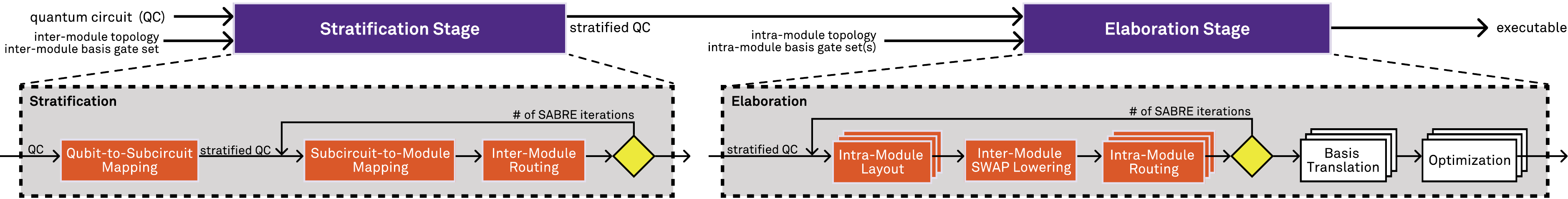}
    \caption{High-level diagram of SEQC. Elements in \textcolor{nudarkred}{orange} reflect modifications from existing compiler designs.}
    \label{fig:high_level_diagram}
\end{figure*}

SEQC incorporates the hierarchical nature of modular architectures by adopting a two-stage procedure (\cref{fig:high_level_diagram}): a one-time \textbf{stratification stage} that elevates a quantum circuit to the module level, and a per-module \textbf{elaboration stage} that can be repeated cheaply to recompile the quantum circuit with the most up-to-date backend properties.

\subsubsection{Stratification Stage}
\begin{figure*}[tbp]
    \centering
    \hfill
    \begin{subfigure}[t]{0.22\textwidth}
        \centering
        \includegraphics[width=0.64\textwidth]{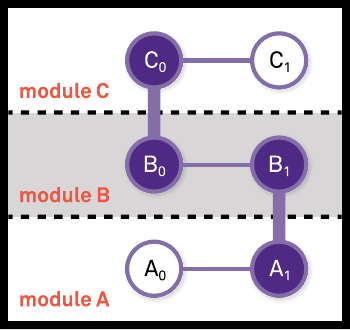}
        \caption{Example module topology}
        \label{fig:stratify_topology}
    \end{subfigure}
    \hfill
    \begin{subfigure}[t]{0.24\textwidth}
        \centering
        \includegraphics[width=\textwidth]{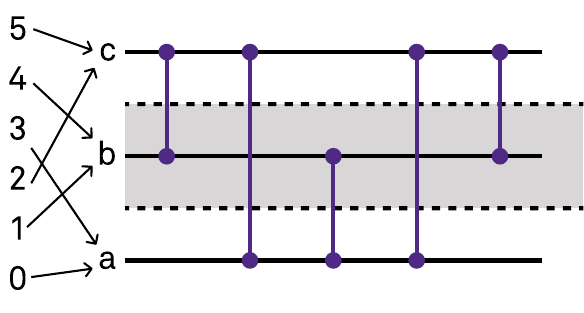}
        \caption{Qubit-to-subcircuit mapping}
        \label{fig:stratify_qubit_to_subcircuit}
    \end{subfigure}
    \hfill
    \begin{subfigure}[t]{0.24\textwidth}
        \centering
        \includegraphics[width=\textwidth]{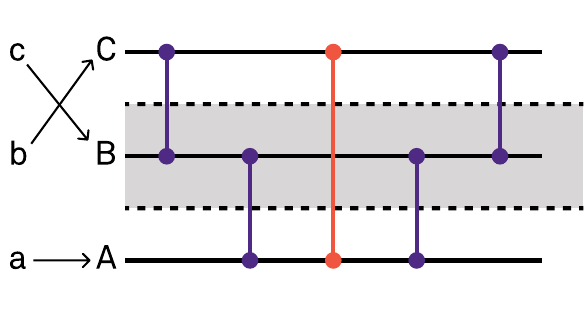}
        \caption{Subcircuit-to-module routing}
        \label{fig:stratify_subcircuit_to_chiplet}
    \end{subfigure}
    \hfill
    \begin{subfigure}[t]{0.24\textwidth}
        \centering
        \includegraphics[width=\textwidth]{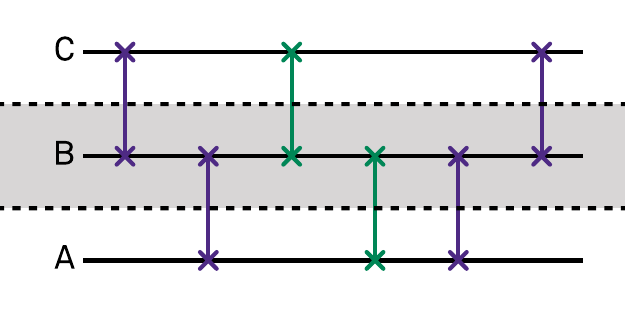}
        \caption{Inter-module routing}
        \label{fig:stratify_inter_chiplet_routing}
    \end{subfigure}
    \hfill
    \vspace{-0.05in}
\caption{Step-by-step example of inter-module compilation in the stratification stage. Numbers denote qubits; lowercase letters denote subcircuits, while uppercase letters denotes physical modules.}
\label{fig:stratify}
\end{figure*}

In this stage, we map a quantum circuit to the inter-module device topology (\cref{fig:stratify_topology} in our running example) by stratifying it into smaller, interconnected \emph{subcircuits} (\cref{fig:stratify_qubit_to_subcircuit,fig:stratify_subcircuit_to_chiplet,fig:stratify_inter_chiplet_routing}).
First, \textbf{qubit-to-subcircuit mapping} maps each virtual qubit to a virtual subcircuit. Then, \textbf{module allocation} maps the virtual subcircuits to physical modules on the target device and routes inter-module gates to conform with module connectivity.
\paragraph{Qubit-to-Subcircuit Mapping}\label{sec:q2s-mapping}

Qubit-to-subcircuit mapping produces a \emph{stratified} quantum circuit, where virtual qubits (numbers in \cref{fig:stratify_qubit_to_subcircuit}) are grouped into subcircuits (lowercase letters). 
Minimizing the number of connections between subcircuits in the mapping thereby reduces inter-module SWAPs between modules. We investigate two methods for this tasks: Simulated Annealing and Spatio-Temporal Aware Qubit Allocation (STA)~\cite{ovide2024sta}.
Our simulated annealing method 
(\cref{alg:SA-q2s-mapping})
starts with an initial qubit-to-subcircuit mapping, which is then refined to minimize a cost function defined 
by the number of inter-module gates given a certain layout. 
A ``kick'' function is then applied iteratively to swap qubits in the layout to try to gather strongly interacting qubits and minimize the number of inter-module gates. The algorithm accepts all improving kicks, but also admits worsening ones with a gradually decreasing probability in hope of escaping local minima.
\begin{algorithm}[t]
    \caption{Qubit-to-Subcircuit Mapping---Sim. Annealing}
    \label{alg:SA-q2s-mapping}

    \small
    \hspace*{\algorithmicindent}\textbf{Input:} input circuit $C$; module count $N_C$; module size $N_Q$\\
    \hspace*{\algorithmicindent}\textbf{Const:} starting temp. $T_0$; kicks per degree $\alpha$; cooling factor $\beta$
    \begin{algorithmic}[1]
        \Procedure{Cost}{$C, L$}\label{alg:q2s-mapping:cost}
            \State $c \gets 0$ \Comment{Count}
            \For{$g \in \textsc{TwoQubitGates}(C)$}
                \State $q_0,q_1 \gets \textsc{Qubits}(g)$
                
                \State $s_0, s_1 \gets L[q_0], L[q_1]$

                \State Increment $c$ \textbf{if} $s_0 \ne s_1$
            \EndFor
            \State \Return $c$
        \EndProcedure
        \Procedure{Kick}{$D, N_C, N_Q, L$}
            \State $L' \gets L$ \Comment{Copy layout}
            \For{$i = 1,\dots,D$}
                \State $s_0 \gets \textsc{RandChoice}(\{1,\dots,N_C\})$
                \State $s_1 \gets \textsc{RandChoice}(\{1,\dots,N_C\} \backslash \{s_0\})$
                \State $q_0,q_1 \gets \textsc{RandChoice}(L'[s_0]), \textsc{RandChoice}(L'[s_1])$
                \State $L'[s_0] \gets (L'[s_0] \cup \{q_1\}) \backslash \{q_0\}$ 
                \State $L'[s_1] \gets (L'[s_1] \cup \{q_0\}) \backslash \{q_1\}$ \Comment{Swap $q_0$ and $q_1$}
            \EndFor
            \State \Return $L'$
        \EndProcedure
        \Procedure{Q2SMappingAN}{$C, N_C, N_Q, T_0, \alpha, \beta$}
            \State $L \gets $ Initial Layout  \Comment{Current Solution}
            \State $R \gets L$ \Comment{Best Solution}
            \State $T \gets T_0$ \Comment{Annealing Temperature}
            \While{$T > 1$}
                \State $D \gets \textsc{Ceil}(T \times \alpha)$ \Comment{Number of kicks to make}
                \State $L' \gets \textsc{Kick}(D, N_C, N_Q, L)$
                \State $\Delta \gets \textsc{Cost}(C, L') - \textsc{Cost}(C, L)$
                \State $k \gets \textsc{Rand}(0, 1)$ \Comment{Random number in $[0, 1) \subset\mathbb{R}$}
                \State $L \gets L'$ \textbf{if} $(\Delta < 0)$ or $(k > \exp(-\Delta / T))$\label{alg:q2s-mapping:acceptance-func}
                \State $R \gets L'$ \textbf{if} $(\textsc{Cost}(C, L') < \textsc{Cost}(C, R))$
                \State $T \gets T \times \beta$ \Comment{Decrease temperature}
            \EndWhile
            \State \Return $R$
        \EndProcedure
    \end{algorithmic}
    \normalsize
\end{algorithm}
STA is a qubit allocation algorithm for trapped-ion quantum systems. These systems are often modular and composed of traps, each containing multiple qubits (ions).
STA tries to minimize overheads due to inter-trap associated qubit shuttling and relocation~\cite{ovide2024sta}, not unlike our incentive to minimize the number of inter-module interactions. 
Thus, we adapted a component of the STA algorithm (Algorithm 2 of Ovide \textit{et al.}~\cite{ovide2024sta}) 
to perform qubit-to-subcircuit mapping and generate an initial subcircuit-to-module mapping
(\cref{alg:sta-q2s-mapping}).

Like the original STA algorithm, we prioritize mapping qubits that frequently interact with other qubits, or have a large degree of direct and indirect interactions between every qubit pair. This information, quantified by $R$ and $T$ in \cref{alg:sta-q2s-mapping}, is then used for qubit allocation.
Then, for each iteration of \cref{alg:sta-q2s-mapping}, a pair of qubits $(q_1, q_2)$ is extracted from $T$ and the path along $(q_1, q_2)$ is routed. 
STA also recursively maps any qubits more strongly connected to $q_2$ than $q_1$ to keep strongly connected qubits physically close together.

\begin{algorithm}[t]
    \caption{Qubit-to-Subcircuit Mapping with STA}
    \label{alg:sta-q2s-mapping}
    \small
    \begin{algorithmic}[1]
        \Procedure{ComputeRatios}{$C$}
            \State $R \gets $ empty map \Comment{ordered in descending values}
            \For{$q \in \textsc{Qubits}(C)$}
                \State $R[q] \gets \frac{\textsc{InteractionCnt}(q, C)} {\textsc{NumQubits}(C)}$ 
                \Comment{fraction of qubits that interacts with $q$}
            \EndFor
            \State \Return $R$
        \EndProcedure
        \Procedure{ComputeST}{$C, S$}
            \State $T \gets$ empty map \Comment{ordered in descending values}
            \For{$(q_1, q_2) \in \textsc{AllPairs}(\textsc{Qubits}(C))$}
                \For{$(i,s) \in \textsc{enumerate}(S)$}
                    \State Add $2^{-i}$ to $T[(q_1, q_2)]$ \textbf{if} $q_1$ and $q_2$ interact in $s$
                \EndFor
            \EndFor    
            \State \Return $T$
        \EndProcedure
        \Procedure{MapQubit}{$q_1, T, R, L$}
            \State $P \gets$ first pair in $T$ containing $q_1$; $q_2 \gets$ other qubit in $P$
            \If{$\textsc{Index}_R(q_2) < \textsc{Index}_R(q_1)$} \Comment{  if $q_2$ interacts more}
                \State 
                $\textsc{MapQubit}(q_2, T, R, L)$ 
            \EndIf
            \If {$q_1 \in R$ or $q_2 \in R$}
                 \State $\textsc{PlaceQubits}(q_1, q_2, L)$
                 \State $R.\textsc{remove}(q_1, q_2)$; $T.\textsc{remove}(P)$ 
            \EndIf
        \EndProcedure
        \Procedure{Q2SMappingSTA}{$C$} \Comment{$C$: Quantum Circuit}
            \State $S \gets \textsc{Layers}(C)$; $L \gets$ initial layout
            \State $R \gets \textsc{ComputeRatios}(C)$; $T \gets \textsc{ComputeST}(C, S)$
            \While{$R$ is not empty}
                \State $\textsc{MapQubit}(R[0], T, R, L)$
            \EndWhile
            \State \Return $L$
        \EndProcedure
    \end{algorithmic}
    \normalsize
\end{algorithm}

\paragraph{Module Allocation} \label{sec:module-alloc}

\begin{algorithm}[t]
    \caption{Inter-Module Routing}
    \label{alg:paul}
    \small
    \begin{algorithmic}[1]
        \Procedure{FindSwapCandidates}{$F, E$}
            \State \Return $\textsc{FindSymbioticSwaps}(F, E)$ \textbf{if} non-empty
            \State \Return $\textsc{FindCommensalisticSwaps}(F, E)$ \textbf{if} non-empty
            \State \Return $\textsc{FindParasiticSwaps}(F, E)$
        \EndProcedure
        \Procedure{IntraModuleGates}{$G, L$}
            \State \Return $\{g : g \in G~\wedge~ \left|\{L[q]:q \in \textsc{Qubits}(g)\}\right| = 1\}$
        \EndProcedure
        \Procedure{InterModuleGates}{$G, L$}
            \State \Return $\{g : g \in G~\wedge~ \left|\{L[q]:q \in \textsc{Qubits}(g)\}\right| > 1\}$
        \EndProcedure
        \Procedure{AdvanceFrontier}{$C_\text{out}, D, L$}
            \State $F \gets \textsc{FirstLayer}(D)$
            \For{$g \in \textsc{IntraModuleGates}(F, L)$}
            \State $C_\text{out}.\textsc{append}(g)$
            \State $D.\textsc{remove}(g)$
            \EndFor
            \State \Return $C_\text{out}, D$
        \EndProcedure
        \Procedure{Heuristic}{$D, L, s$}
            \State $D', L' \gets \textsc{ApplySwap}(D, L, s)$
            \State $H \gets \textsc{InterModuleGates}(\textsc{Gates}(D'), L')$
            \State \Return $|H|$
        \EndProcedure
        \Procedure{InterModuleRouting}{$C, L, T$}
        \LComment{$C$: quantum circuit, $L$: qubit-to-subcircuit mapping from \cref{alg:sta-q2s-mapping}, $T$: the module topology.}
            \State $E \gets$ All-pair distances of module topology $T$
            \State $C_\text{out} \gets \text{empty circuit}$ \Comment{Output Circuit}
            \State $D \gets \textsc{CircuitToDAG}(C)$
            \State $C_\text{out}, D \gets \textsc{AdvanceFrontier}(C_\text{out}, D, L)$
            \While{$D$ is not empty}
                \State $\mathcal{S} \gets \textsc{FindSwapCandidates}(\textsc{FirstLayer}(D), E)$
                \State $s^* \gets \argmin_{s \in \mathcal{S}} \textsc{Heuristic}(D, L, s)$
                \State $D, L \gets \textsc{ApplySwap}(D, L, s^*)$
                \State $C_\text{out}$.\textsc{append}($s^*$)
                \State $C_\text{out}, D \gets \textsc{AdvanceFrontier}(C_\text{out}, D, L)$
            \EndWhile
            \State \Return $C_\text{out}$
        \EndProcedure
    \end{algorithmic}
    \normalsize
\end{algorithm}
Analogous to qubit layout and routing, in \cref{fig:high_level_diagram} we divide module allocation into  \emph{subcircuit-to-module mapping} (\cref{fig:stratify_subcircuit_to_chiplet}), an initial assignment of subcircuits to physical modules, and \emph{inter-module routing} (\cref{fig:stratify_inter_chiplet_routing}), which inserts inter-module SWAPs to account for limited inter-module gate connectivity. 
Note that these inter-module SWAPs operate at the abstract module level, and concrete link assignment occurs in a later stage.
In our design, we use our qubit-to-subcircuit mapping to provide an initial subcircuit-to-module mapping. Then, we extend and modify the SABRE algorithm~\cite{sabre_swap} to apply to modular architectures, placing iterations of inter-module routing alongside randomized qubit permutations to inform a final subcircuit-to-module mapping.%

Inter-module routing seeks to consolidate the qubits of non-SWAP inter-module gates to a single module by inserting inter-module SWAPs into the stratified circuit. Like qubit routing, as a stratified circuit contains multiple inter-module gates, we would expect it to be impractical to derive an optimal solution for inter-module routing at compile-time. Thus, we employ a similar SWAP selection scheme as SABRE~\cite{sabre_swap}, where the inserted SWAPs are chosen from a pool of SWAP candidates according to a heuristic module distance score.

To elaborate, we categorize SWAP candidates into three tiers---\emph{symbiotic}, \emph{commensalistic}, and \emph{parasitic}---based on their impact to the inter-module gates, as shown in \cref{alg:paul}. 
The three tiers are ordered from most to least beneficial to inter-module routing.
\emph{Symbiotic} SWAPs directly benefit routing inter-module gates. This is when each qubit of the SWAP corresponds to a different inter-module gate, and inserting the SWAP reduces the module distance score for both gates. Symbiotic SWAPs are of the highest priority as they represent the most efficient possible form of SWAP insertion.
The next tier, \emph{commensalistic} SWAPs do not increase the module distance of any directly impacted gate. This typically occurs when an idle qubit serves as an ancilla qubit to facilitate routing an inter-module gate without affecting other inter-module gates. In certain device topologies, a SWAP acting on the qubits of two inter-module gates can also exhibit commensalistic behavior.
Finally, \emph{parasitic} SWAPs act on the qubits of two inter-module gates, where one gate benefits while the other is harmed. As they are actively counterproductive towards one of the gates, these SWAPs will be avoided in most cases, only being necessary in hard routing instances or deadlocks.

\subsubsection{Elaboration Stage}

\begin{figure*}[tbp]
    \centering
    \begin{subfigure}[t]{12em} 
        \centering
        \includegraphics[width=5.5em]{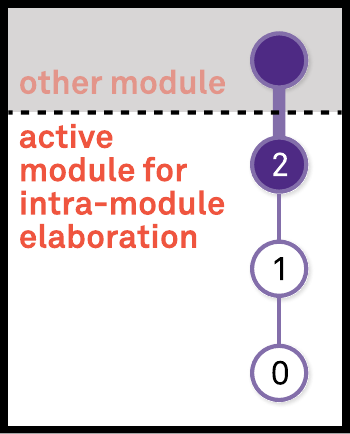}
        \captionsetup{justification=centering, singlelinecheck=false}
        \caption{Intra-module topology\\example}
        \label{fig_elaborate_topology}
    \end{subfigure}
    \hfill
    \begin{subfigure}[t]{10.5em} 
        \centering
        \includegraphics[width=\textwidth]{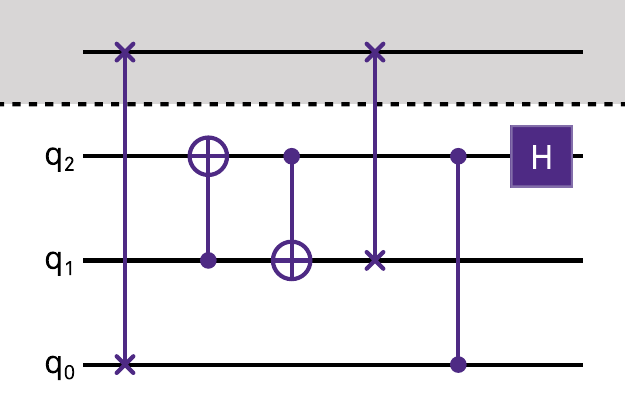}
        \captionsetup{justification=centering, singlelinecheck=false}
        \caption{Stratified 3-qubit\\subcircuit}
        \label{fig:elaborate_initial}
    \end{subfigure}
    \hfill
    \begin{subfigure}[t]{14em} 
        \centering
        \includegraphics[width=9em]{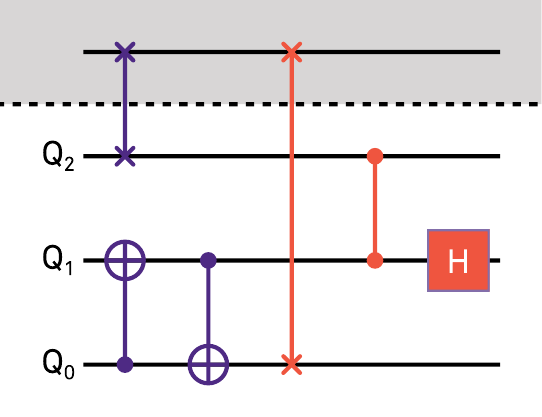}
        \captionsetup{justification=centering, singlelinecheck=false}
        \caption{Subcircuit after intra-module layout mapping $[q_0,q_1,q_2]\Rightarrow[Q_2,Q_0,Q_1]$}
        \label{fig:elaborate_layout}
    \end{subfigure}

    \vspace{0.1in}

    \centering
    \begin{subfigure}[t]{12em} 
        \centering
        \includegraphics[width=9em]{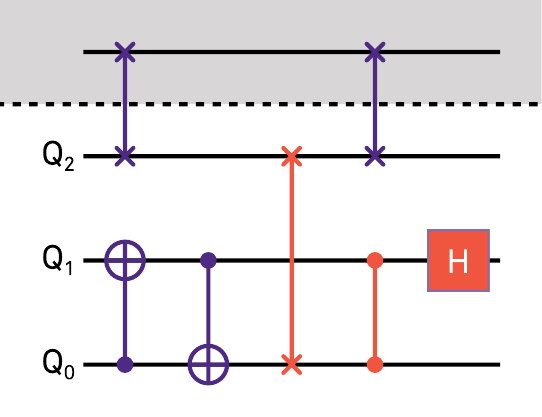}
        \caption{Inter-module SWAP lowering}
        \label{fig:elaborate_nearest}
    \end{subfigure}
    \hfill
    \begin{subfigure}[t]{10.5em} 
        \centering
        \includegraphics[width=\textwidth]{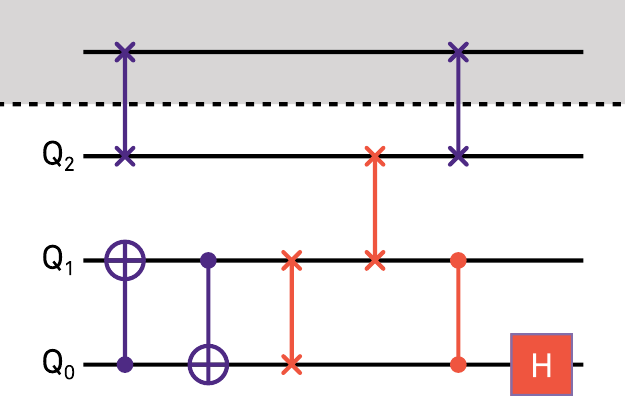}
        \caption{Intra-module routing}
        \label{fig:elaborate_routing}
    \end{subfigure}
    \hfill
    \begin{subfigure}[t]{14em} 
        \centering
        \includegraphics[width=9em]{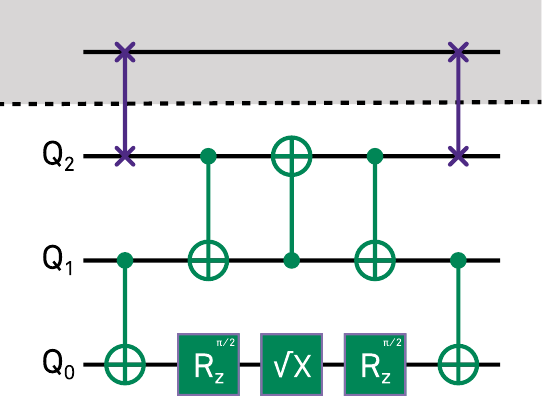}
        \caption{Basis translation and optimization}
        \label{fig:elaborate_translation}
    \end{subfigure}
    \vspace{-0.05in}
\caption{Step-by-step example of compilation in the Elaboration stage. Gates unimplementable on hardware are in \textcolor{nubrightred}{orange}.}
\label{fig_elaborate}
\vspace{-0.1in}
\end{figure*}

Following the stratification stage, the compilation process within each module is largely independent and identical to that of a monolithic quantum architecture. 
The only exceptions are the inter-module SWAPs, which must be assigned to concrete inter-module links at the qubit level, 
and must remain fixed and immutable to subsequent compilation stages, which we refer to as \emph{pinned}. Despite this limitation, restricting the task of compilation to smaller subsets of qubits facilitates parallelization and reduces problem size for the computationally intensive routing and optimization.
Thus, we divide the elaboration stage into two substages: \textbf{qubit allocation} and \textbf{parallel basis translation and optimization}. \cref{fig_elaborate} shows an example of elaboration for a single module (\cref{fig_elaborate_topology}) and its corresponding subcircuit (\cref{fig:elaborate_initial}).

\paragraph{Qubit Allocation}
We employ a modified version of SABRE to perform qubit allocation in the elaboration stage. Specifically, we parallelize the SABRE routing algorithm and add constraints to maintain correctness when applied to a stratified quantum circuit with inter-module gates.%

As noted earlier, the only non-parallelizable task during elaboration is the placement of inter-module gates.
To account for this, we introduce a serial stage (\cref{fig:elaborate_nearest}) between the parallel layout and routing passes that greedily places gates on the nearest valid inter-module edge. 
We choose a greedy policy because inter-module gates are more restrictive to place and experience substantially higher error rates ~\cite{smith_2022_chiplets}. These gates are then \emph{pinned}, imposing an additional constraint on qubit routing. 
To forbid SABRE from modifying inter-module SWAPs, we specialize the gate sets permitted to operated across modules to only inter-module SWAPs, forcing SABRE to error out if it attempts to place illegal gates. In addition, we extend SABRE's cost heuristic to weigh the low-fidelity couplings with higher costs, discouraging it from placing illegal gates on the higher-error inter-module links.

\paragraph{Parallel Translation and Optimization} \label{sec:design_solve_optimization}

The remaining compilation stages of basis translation and circuit optimization can be easily parallelized at the module granularity (\cref{fig:elaborate_translation}). Basis translation, in particular, is embarrassingly parallel.
Meanwhile, as discussed in \cref{sec:background}, optimization algorithms already employ partitioning and parallelization to create tractable subproblems from intractably large circuits.
Partitioning through natural module boundaries enables us
to reap the same resource and performance benefits of partitioning without compromising on solution quality.

\section{Methodology}
We evaluate quantum compilers based on the their solution quality, compilation speed, and scalability. Each compiler is benchmarked
on a suite of quantum circuits, detailed in \cref{sec:benchmarks}, targeting mock modular quantum backends as specified in \cref{sec:device_specifications}.
We evaluate four compilers---a reference compiler, a peephole-augmented compiler, and two versions of SEQC---on simulated backends ranging from 1 to 100 modules. All compilers are implemented in the Qiskit SDK~\cite{qiskit2024} v$2.3.0$ and Python $3.12.6$, and timing results are taken on a machine with a $128$-core Ampere Altra Max ARM64 processor and $256$ GB of DDR4 RAM. 

Our reference compiler is the Qiskit builtin compiler (with \texttt{optimization\_level=3}). As it is not module-aware, we run these experiments on a backend with \emph{universal} inter-module links (\cref{sec:device_specifications}).
The peephole-augmented Qiskit compiler is modified from Qiskit by inserting our peephole pass between its \texttt{routing} and \texttt{basis\_translation} stages, thus making it module-aware. 
Our two SEQC versions differ in the qubit-to-subcircuit mapping techniques. The first version leverages simulated annealing, initialized with a starting temperature $T_0 = 200$, a rate of change of $0.005T$ per round, $\frac{T}{50}$ permutations per round, and a total of $10$ trials per CPU core. The second version leverages STA for qubit-to-subcircuit and subcircuit-to-module mapping.

We aggregate results in a manner similar to SPEC benchmarks~\cite{spec2017}, by taking the geometric mean of the relative performance ratio against a baseline technique across a benchmark suite of quantum circuits.
The resultant score is then derived for each of the evaluated metrics on multiple backends of varying numbers of modules.

\subsection{Benchmark Suite} \label{sec:benchmarks}
We leverage circuits from \texttt{Supermarq}~\cite{supermarq} that are representative of practical, real-world applications for quantum computing, while also being feasible to compile for a large number of qubits.
In particular, our suite comprises the following quantum circuits. 

\begin{figure*}[t]
    \centering
    \hfill
    \begin{subfigure}[t]{0.14\linewidth}
        \centering
        \includegraphics[height=4.5em]{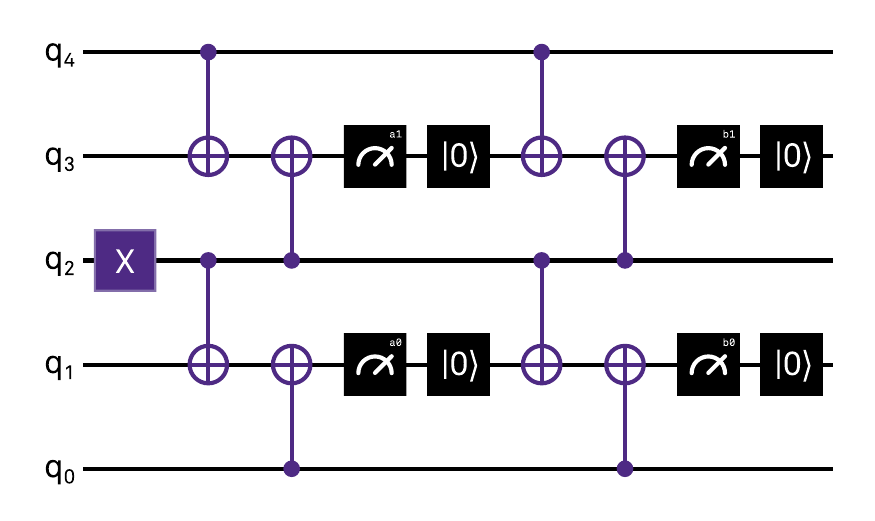}   
        \caption{BitCode}
        \label{fig:benchmark-circ_bitcode}
    \end{subfigure}
    \hfill
    \begin{subfigure}[t]{0.23\linewidth}
        \centering
        \includegraphics[height=4.5em]{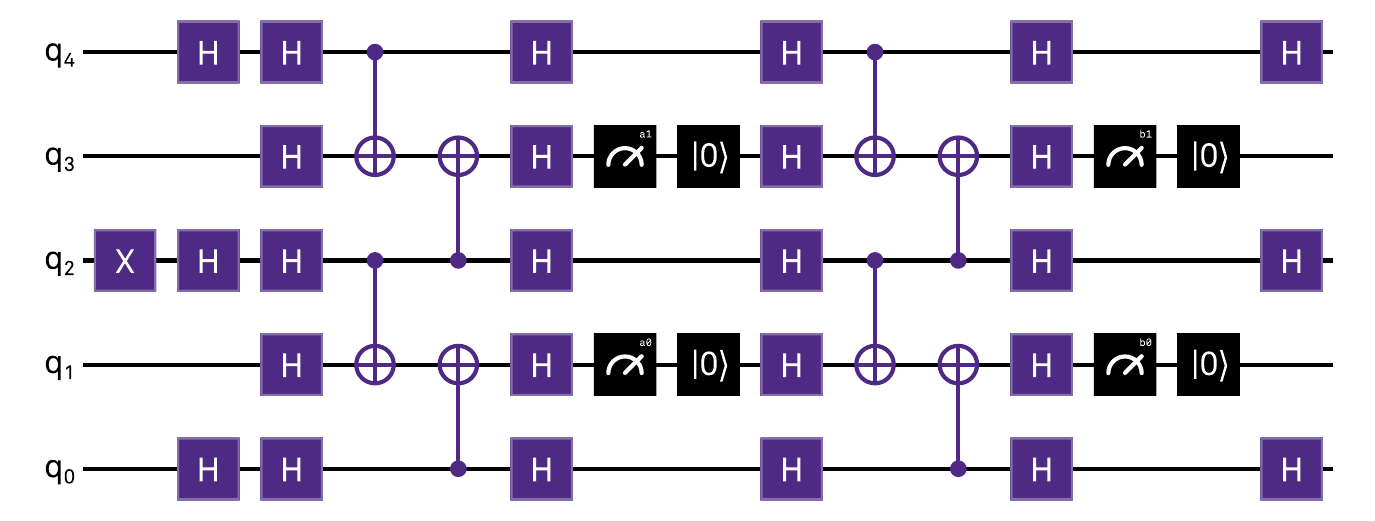}   
        \caption{PhaseCode}
        \label{fig:benchmark-circ_phasecode}
    \end{subfigure}
    \hfill
    \begin{subfigure}[t]{0.09\linewidth}
        \centering
        \includegraphics[height=4.5em]{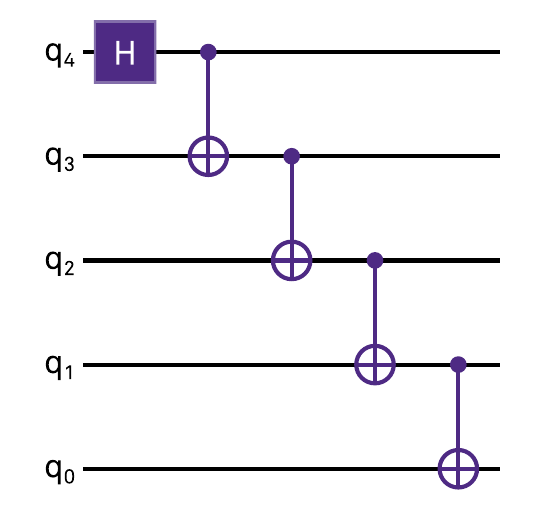}   
        \caption{GHZ}
        \label{fig:benchmark-circ_GHZ}
    \end{subfigure}
    \hfill
    \begin{subfigure}[t]{0.24\linewidth}
        \centering
        \includegraphics[height=4.5em]{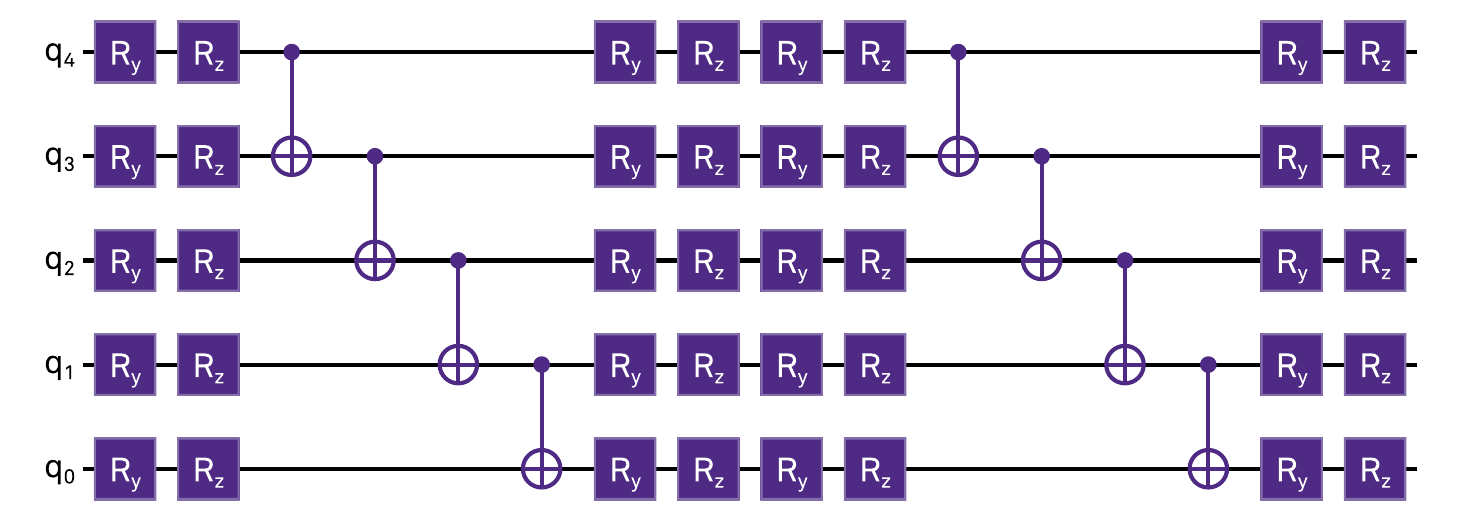}   
        \caption{VQE}
        \label{fig:benchmark-circ_VQE}
    \end{subfigure}
    \hfill
    \begin{subfigure}[t]{0.23\linewidth}
        \centering
        \includegraphics[height=4.5em]{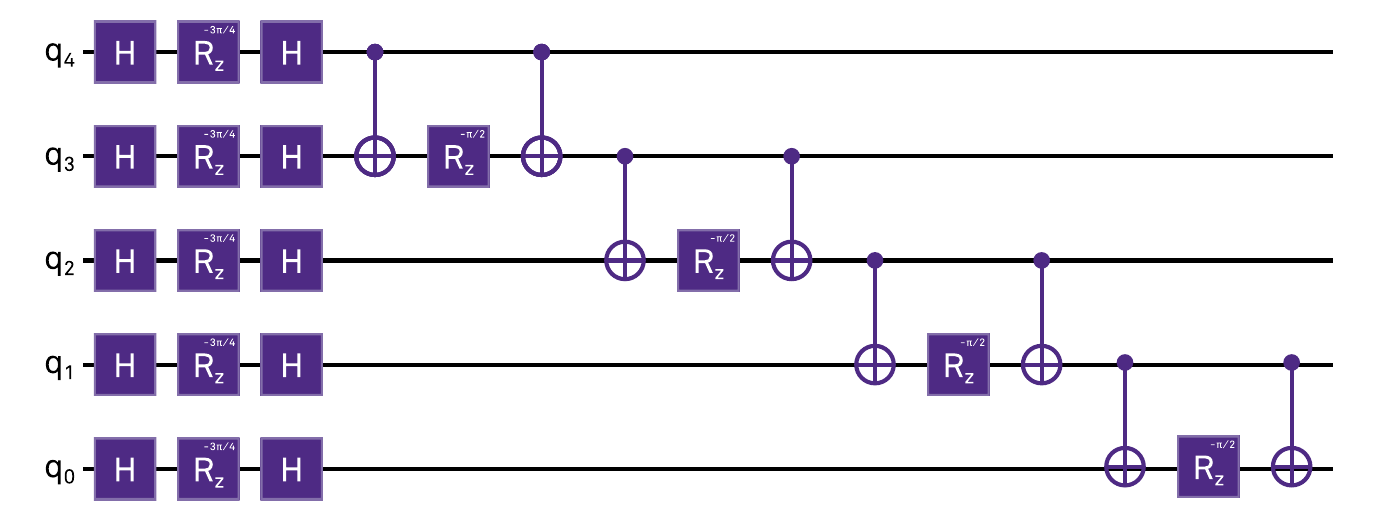}   
        \caption{Hamiltonian Sim.}
        \label{fig:benchmark-circ_hamiltonian}
    \end{subfigure}
    \hfill
    \vspace{-0.05in}
    \caption{Examples of benchmark applications for 5-qubit circuits.}
    \label{fig:benchmark-circuits}
\end{figure*}

\emph{Bit codes} (\cref{fig:benchmark-circ_bitcode}) and \emph{phase codes} (\cref{fig:benchmark-circ_phasecode}) are essential in quantum error correction. An $n$-qubit benchmark contains $\lfloor \frac{n+1}{2} \rfloor$ data qubits initialized to $\ket{1010...10}$ and two rounds of error correction. These circuits are highly parallelizable and examine a compiler's ability to exploit gate level parallelism. Also, the presence of additional one-qubit gates in phase code circuits may showcase a compiler's ability to differentiate between the relative cost (duration, fidelity) between one- and two-qubit gates.



Next, \emph{Greenberger--Horne--Zeilinger} (GHZ, \cref{fig:benchmark-circ_GHZ}) and \emph{variational quantum eigensolver} (VQE, \cref{fig:benchmark-circ_VQE})~\cite{vqe} primarily consist of long chains of serial CNOT gates, exercising the compiler's qubit allocation. An $n$-qubit CNOT chain that fully utilizes a device with $k$-qubit modules necessitate $\lfloor\frac{n}{k}\rfloor - 1$ CNOT operations to extend between modules. In turn, this will necessitate at least $2\left(\lfloor\frac{n}{k}\rfloor - 1\right)$ inter-module SWAPs to transfer qubits in/out of a module. In particular, we adopt a two-layer VQE ansatz with randomly generated parameters. 



Finally, \emph{Hamiltonian Simulation} benchmarks model the 1D Transverse Field Ising Model system (\cref{fig:benchmark-circ_hamiltonian}). These circuits are highly serial but more complex than the GHZ and VQE circuits, testing the compiler's ability exploit parallelism originating from the efficient placement of SWAP gates.

\subsection{Simulated Quantum Device Specifications} \label{sec:device_specifications}

\begin{figure}[b]
    \centering
    \hfill
    \begin{subfigure}[t]{0.45\columnwidth}
        \centering
        \includegraphics[height=10em]{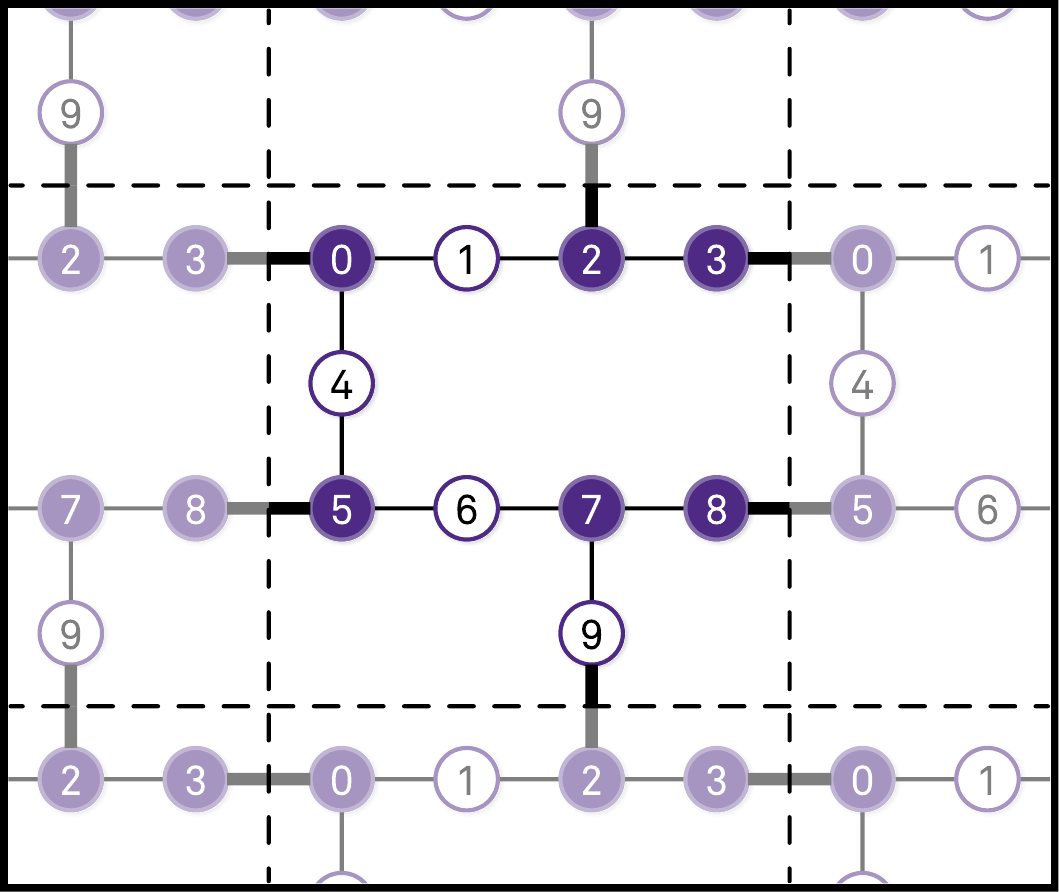}
        \subcaption{
        Small-scale heavy-hex modules, on grid lattice. %
        }
        \label{fig:heavy_hex_topology}
    \end{subfigure}
    \hfill
    \begin{subfigure}[t]{0.45\columnwidth}
    \centering
        \includegraphics[height=10em]{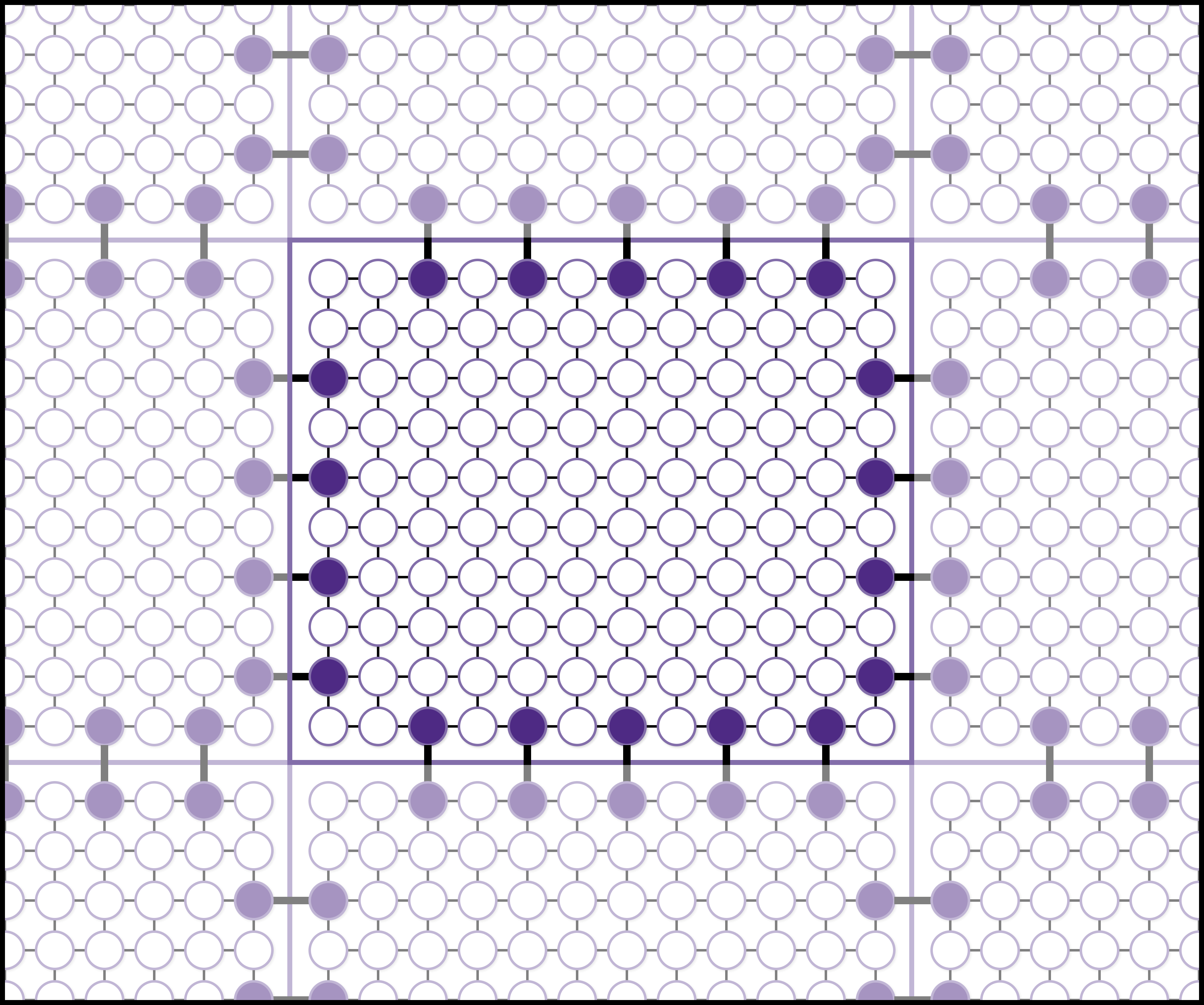}
    \subcaption{Large-scale modules with IBM Nighthawk-like topology.}
    \label{fig:nighthawk_topology}
    \end{subfigure}
    \hfill
    \vspace{-0.05in}
    \caption{Modeled topologies.}
    \label{fig:topologies}
\end{figure}

\begin{table}[b]
    \centering
\begin{subtable}[t]{0.33\columnwidth}
    \centering
    \begin{tabular}{|c||c|}
    \toprule
        T1~\cite{backend_specs} & $20\mu$s \\
        T2~\cite{backend_specs} & $30\mu$s \\
        Freq.~\cite{backend_specs} & $6$GHz \\
    \bottomrule
    \end{tabular}
    \caption{Qubit Properties}
    \label{tab:backend_qubit_properties}
\end{subtable}
\begin{subtable}[t]{0.66\columnwidth}
    \centering
    \begin{tabular}{|c||c|c|}
    \toprule
    Instruction & Time (ns) & Error \\
    \midrule
        X~\cite{backend_specs} & $25.0$ & $0.109 \%$ \\
        SX~\cite{backend_specs}& $25.0$ & $0.109 \%$ \\
        $\text{R}_\text{z}(\phi)$~\cite{backend_specs}& $0.0$ & $0.000 \%$ \\
        CZ (intra-mod.)~\cite{backend_specs} & $34.0$ & $0.605 \%$ \\
        CZ (inter-mod.)$^\dag$ & $126.0$ & $2.420 \%$ \\
        SWAP (inter-mod.)$^\ddag$ & $702.4$ & $10.23\%$ \\
        Reset~\cite{backend_specs} & $500.0$ & $0.186 \%$ \\
        Measure~\cite{backend_specs} & $500.0$ & $0.196 \%$ \\
    \bottomrule
    \end{tabular}
    \captionsetup{justification=centering}
    \caption{
        Instruction Properties \\
        \footnotesize{\textnormal{$^\dag$Universal gate set only. 
        $^\ddag$Non-universal gate set only.}}
    }
    \label{tab:backend_instruction_properties}
\end{subtable}
    \caption{Specifications of Simulated Modular Backend.}
    \label{tab:backend_specs}
\end{table}

We model two types of simulated devices differing in their inter-module topology and module size. To isolate the impact of architecture modularity, we evaluate on symmetrical modules based on a transmon qubit technology.
One variant,
denoted by the \emph{heavy-hex backend}
(\cref{fig:heavy_hex_topology}),
is generated to conform to the modular architecture proposed in Smith \textit{et al.}~\cite{smith_2022_chiplets}, 
and features a 10-qubit, heavy-hexagon topology~\cite{heavy_hex}. 
The other, denoted by the \emph{grid backend}
(\cref{fig:nighthawk_topology}), 
uses modules with a 120-qubit grid topology, based on the upcoming IBM Nighthawk processor~\cite{IBM_2025_FTQC, Nighthawk_topology}.
The hardware specifications we adopted for qubits and intra-module connections (\cref{tab:backend_specs}) 
are sourced from Acharya \textit{et al.}~\cite{backend_specs}. 
To generate the expected fidelity of the inter-module SWAPs, we perform a simulated random benchmarking trial and calculate the error rate of a SWAP gate with these backend specifications. 
Following Smith \textit{et al.}~\cite{smith_2022_chiplets}, we conservatively model \textit{inter}-module SWAPs with $4\times$ the error rate and $4\times$ the duration of an \textit{intra}-module SWAP.
Both simulated devices use a grid lattice topology of symmetric modules, and are generated with the ``most square'' topology (e.g., a $12$-module machine would be 
a $3 \times 4$ grid).


\section{Experimental Results}
We evaluate
the Qiskit reference compiler, the peephole-augmented Qiskit (PA-Qiskit) compiler, and SEQC using simulated annealing (SEQC-AN) and STA (SEQC-STA) on both the 10-qubit heavy-hex and the 120-qubit grid backends.
Note that Qiskit is afforded universal inter-module links, while the remaining three are restricted to only SWAPs across chiplets.
We examine how well each compiler's preferred heuristics translate to metrics of circuit quality, including number of inter-module gates, total gate count, and circuit depth. We also investigate compilation scalability, including recurring and non-recurring compilation time, and memory usage. Relative metrics are normalized over the Qiskit reference compiler.



\begin{figure*}[tb]
    \centering
    \begin{subfigure}[h]{0.45\linewidth}
        \centering
        \includegraphics[width=0.7\linewidth]{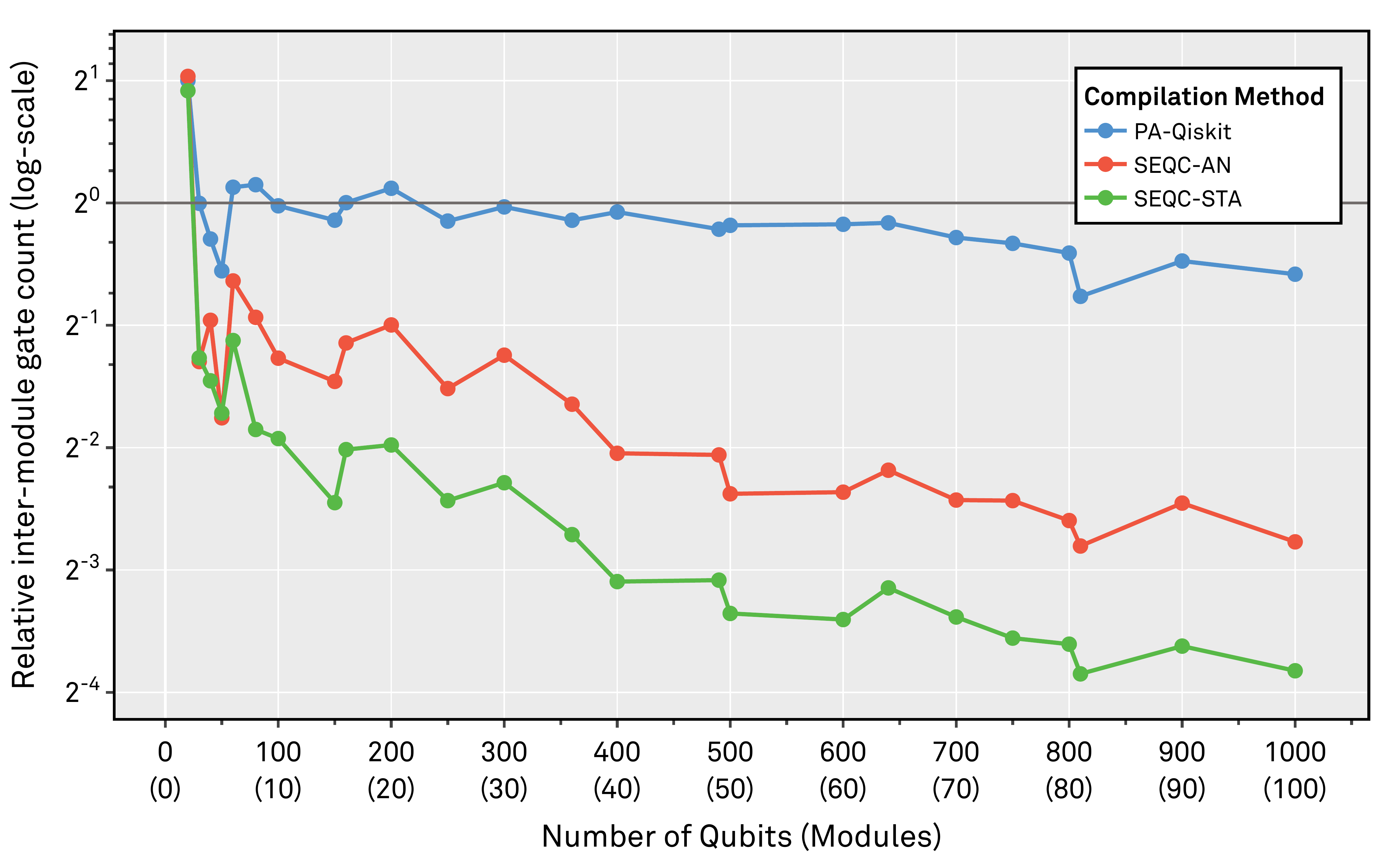}
        \subcaption{Relative inter-module gate count\\(lower is better).}
        \label{fig:10Q_num_inter_chiplet_gates}    
    \end{subfigure}
    \begin{subfigure}[h]{0.45\linewidth}
    \centering
    \includegraphics[width=0.7\linewidth]{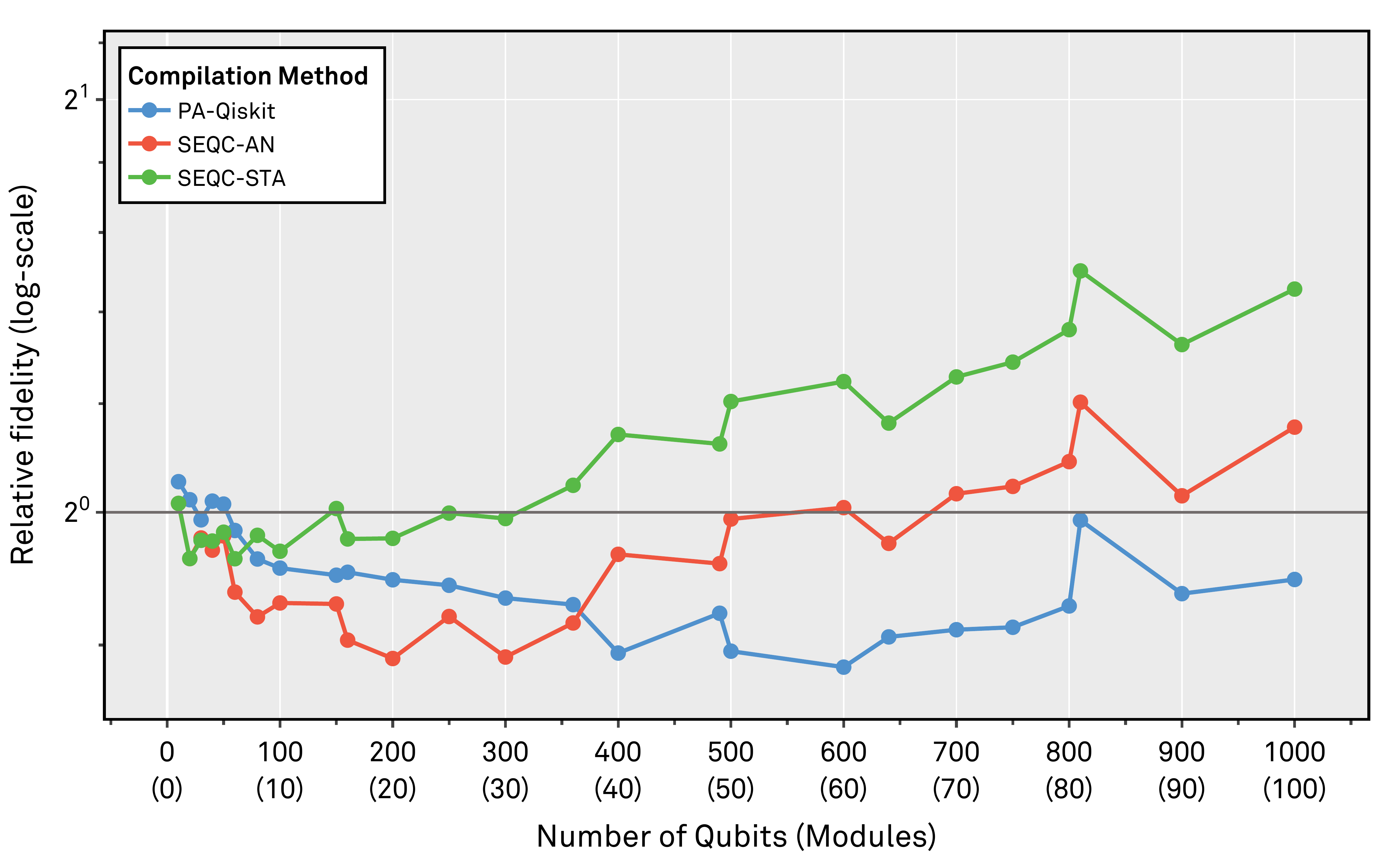}
        \subcaption{Relative estimated fidelity (ESP)\\(higher is better).}
        \label{fig:10Q_fidelity}
    \end{subfigure}
    \vspace{-0.05in}
    \caption{Main circuit performance metrics: inter-module gate count and estimated fidelity, 10-qubit heavy-hex backend.}
    \label{fig:10Q_num_inter_chiplet_gates_and_esp}    
\end{figure*}

\begin{figure*}[tb]
    \centering
    \begin{subfigure}[h]{0.45\linewidth}
        \centering
        \includegraphics[width=0.7\linewidth]{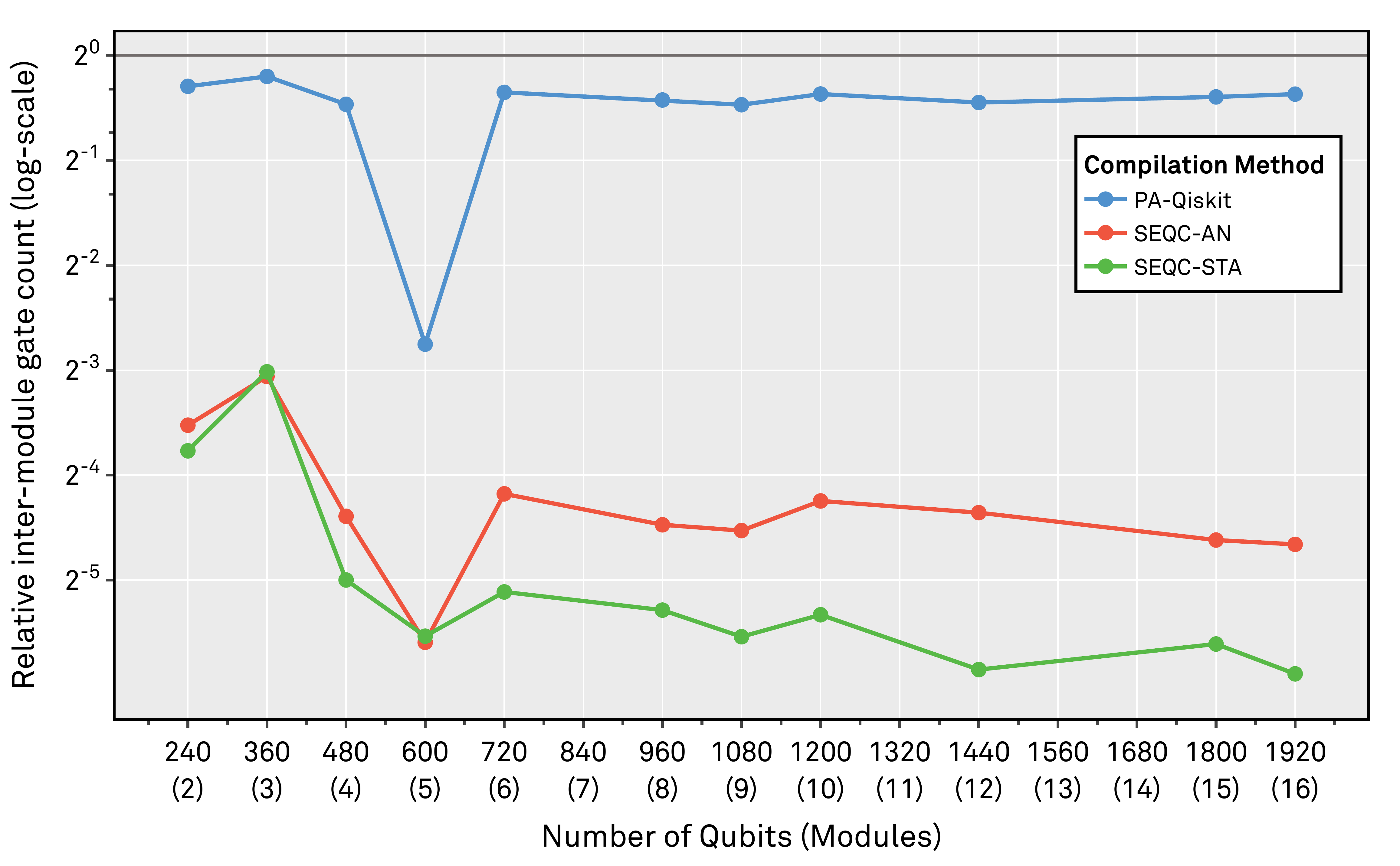}
        \subcaption{Relative inter-module gate count\\(lower is better).}
        \label{fig:120Q_num_inter_chiplet_gates}
    \end{subfigure}
    \begin{subfigure}[h]{0.45\linewidth}
    \centering
        \includegraphics[width=0.7\linewidth]{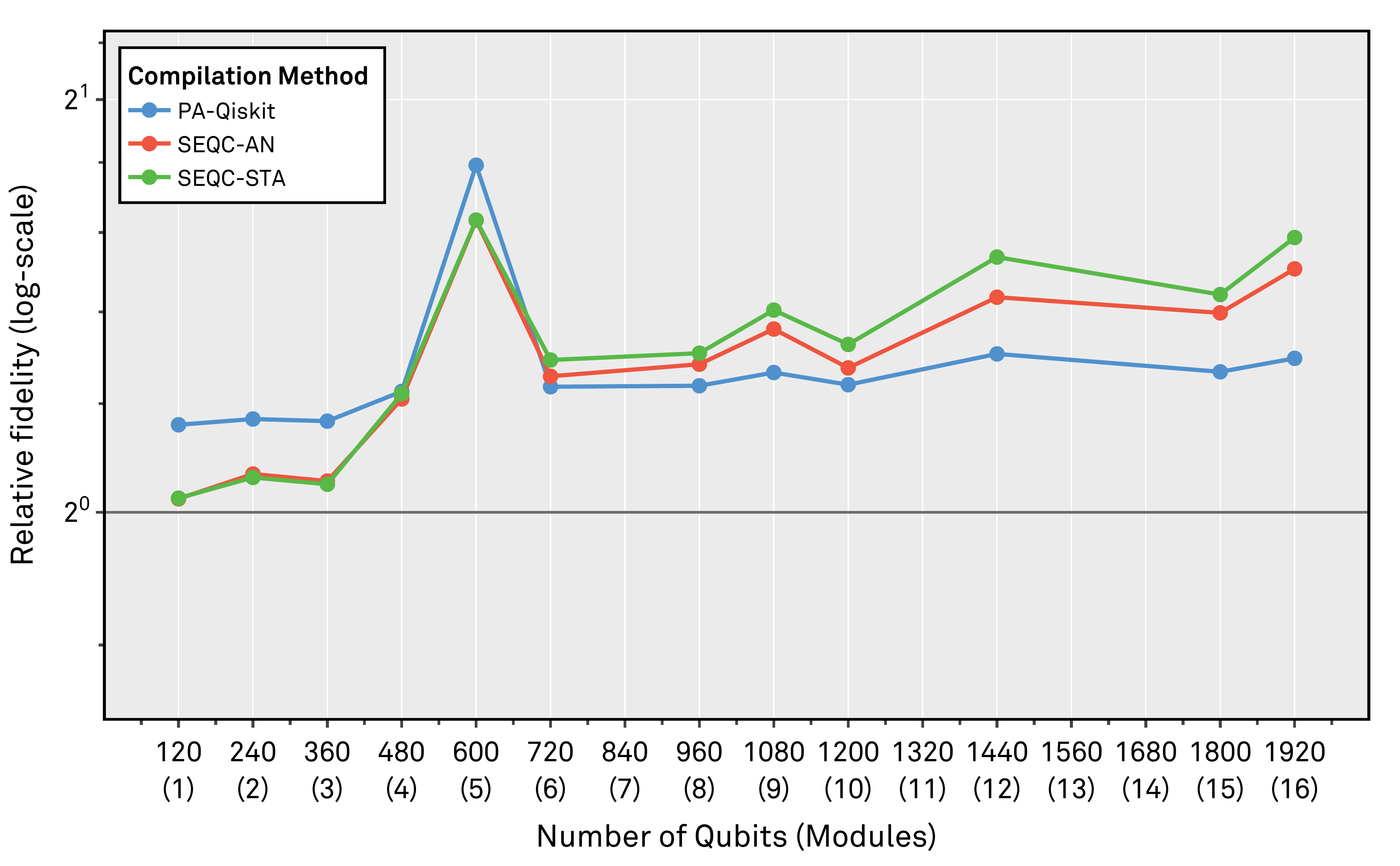}
        \subcaption{Relative estimated fidelity (ESP)\\(higher is better).}
        \label{fig:120Q_fidelity}
    \end{subfigure}
    \vspace{-0.05in}
    \caption{Main circuit performance metrics: inter-module gate count and estimated fidelity, 120-qubit grid backend.}
    \label{fig:120Q_num_inter_chiplet_gates_and_esp}    
\end{figure*}

\paragraph{Number of inter-module gates} 
SEQC primarily minimizes the number of inter-module gates, which are expected to contribute the most to error and execution time relative to other gates. 
Accordingly, SEQC consistently yields significantly fewer inter-module gates compared to Qiskit-based compilers (\cref{fig:10Q_num_inter_chiplet_gates,fig:120Q_num_inter_chiplet_gates}). 
%
Take SEQC-STA for an example. 
On average,
on the heavy-hex~/~grid backends, 
it produces 
$82.43\%~/~96.31\%$ 
fewer inter-module gates than Qiskit,
and up to 
$92.94\%~/~98.32\%$ 
less.
Similarly, compared to PA-Qiskit, it produces 
$64.92\%~/~94.31\%$ 
fewer inter-module gates on average, and up to $78.62\%~/~97.83\%$ less.

For every inter-module gate Qiskit compiles (using the universal backend), we expect PA-Qiskit to produce 2 inter-module SWAP gates. However, we observe PA-Qiskit produces circuits with a roughly comparable number of inter-module gates to Qiskit. In fact, as size increases, PA-Qiskit tends to result in fewer inter-module gates than Qiskit. This outcome may indicate that optimization passes are capable of removing extraneous SWAPs during elaboration.

\paragraph{Fidelity} 

Maximizing fidelity is an essential task for quantum compilers.
We estimate fidelity through estimated success probability (ESP)~\cite{esp}. 
Specifically, we track the per-qubit ESP of a quantum circuit and report an averaged ESP for each benchmark from the geometric mean of the circuit measurements.
\cref{fig:10Q_fidelity,fig:120Q_fidelity} present the relative 
ESP
of each compiler compared to the Qiskit compiler with respect to circuit size.
There is a noticeable benefit in fidelity associated with the SEQC compiler, particularly with SEQC-STA. 
On the heavy-hex~/~grid backend, respectively, we observe an average improvement of $22.13\%~/~3.51\%$ over PA-Qiskit, and up to $62.87\%~/~22.50\%$. 

SEQC similarly outperforms Qiskit by a wide margin---\textbf{notably, even when Qiskit is afforded significant allowances: universal inter-module links and much lower inter-module gate errors} ($2.42\%$ for CZs vs. $10.23\%$ for inter-module SWAPs).
Despite being much more constrained, SEQC-STA demonstrates an average improvement of $9.30\%~/~32.30\%$, and up to $49.99\%~/~63.36\%$, on the heavy-hex~/~grid backend, respectively.

\begin{figure*}[tb]
    \centering
    \begin{subfigure}[t]{0.32\linewidth}
        \centering
        \includegraphics[width=0.99\linewidth]{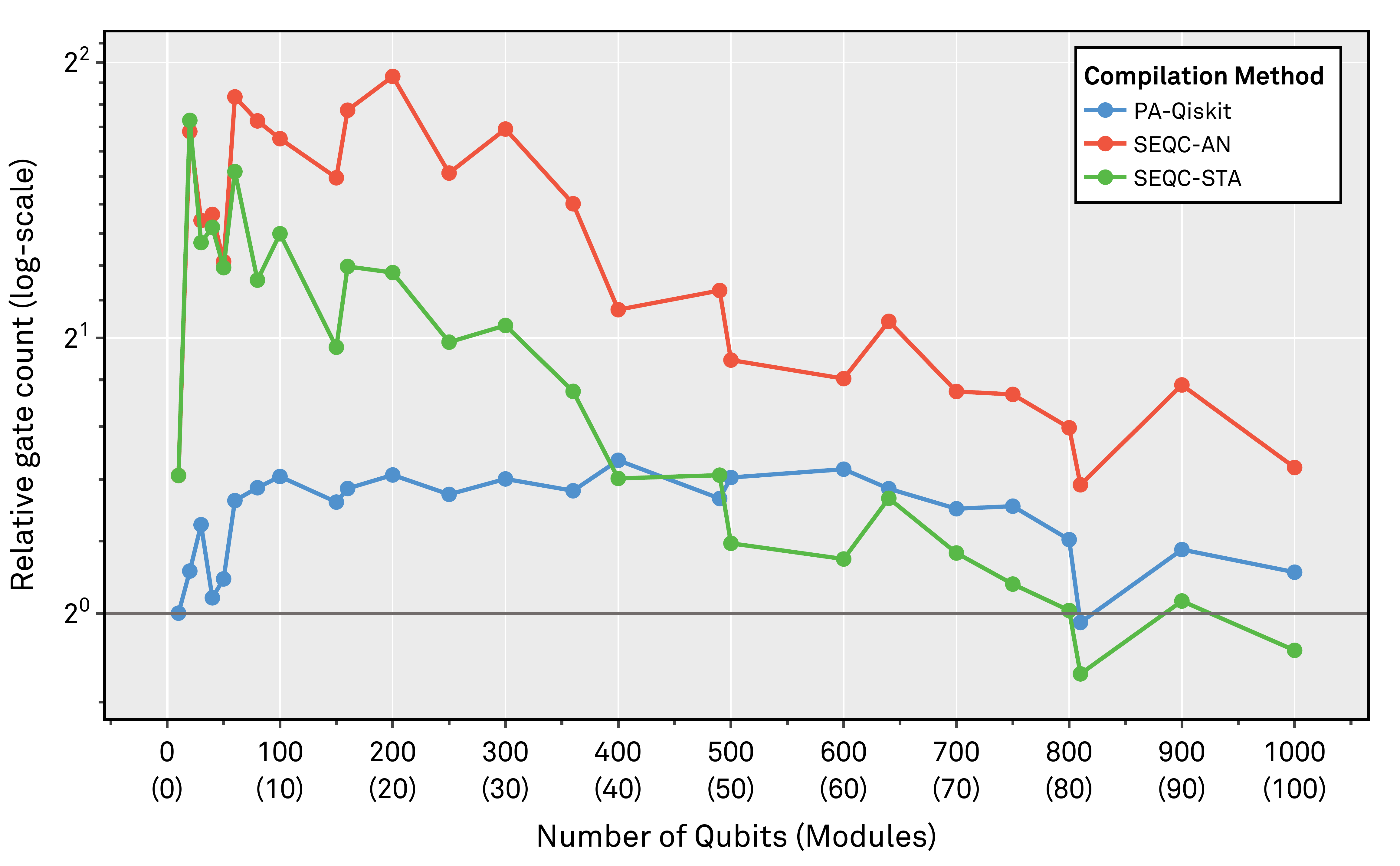}
        \subcaption{Relative total gate count.}
        \label{fig:10Q_gate_count}    
    \end{subfigure}
    \begin{subfigure}[t]{0.32\linewidth}
    \centering
        \includegraphics[width=0.99\linewidth]{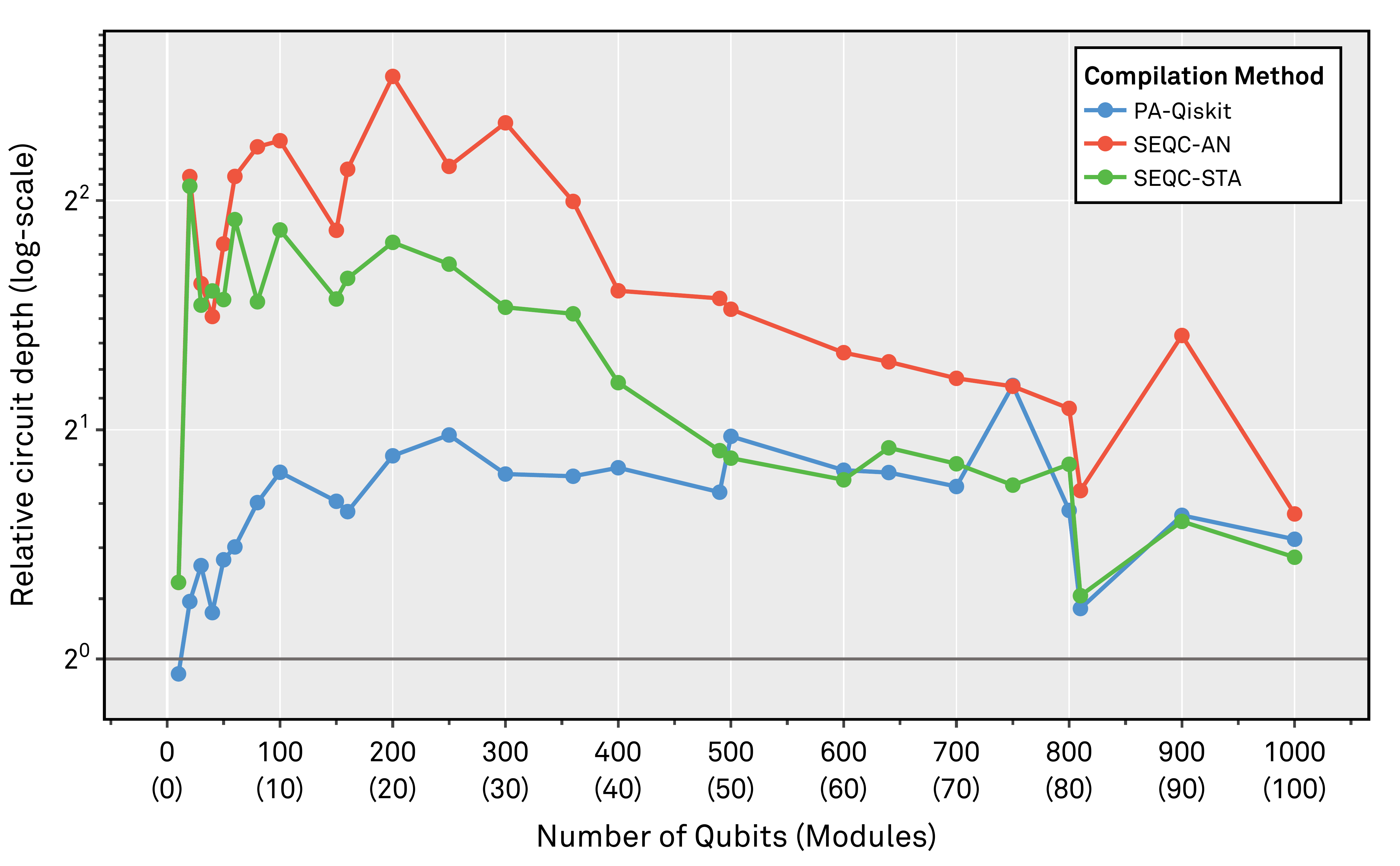}
        \subcaption{Relative circuit depth.}
        \label{fig:10Q_circuit_depth}    
    \end{subfigure}
    \begin{subfigure}[t]{0.32\linewidth}
    \centering
        \includegraphics[width=0.99\linewidth]{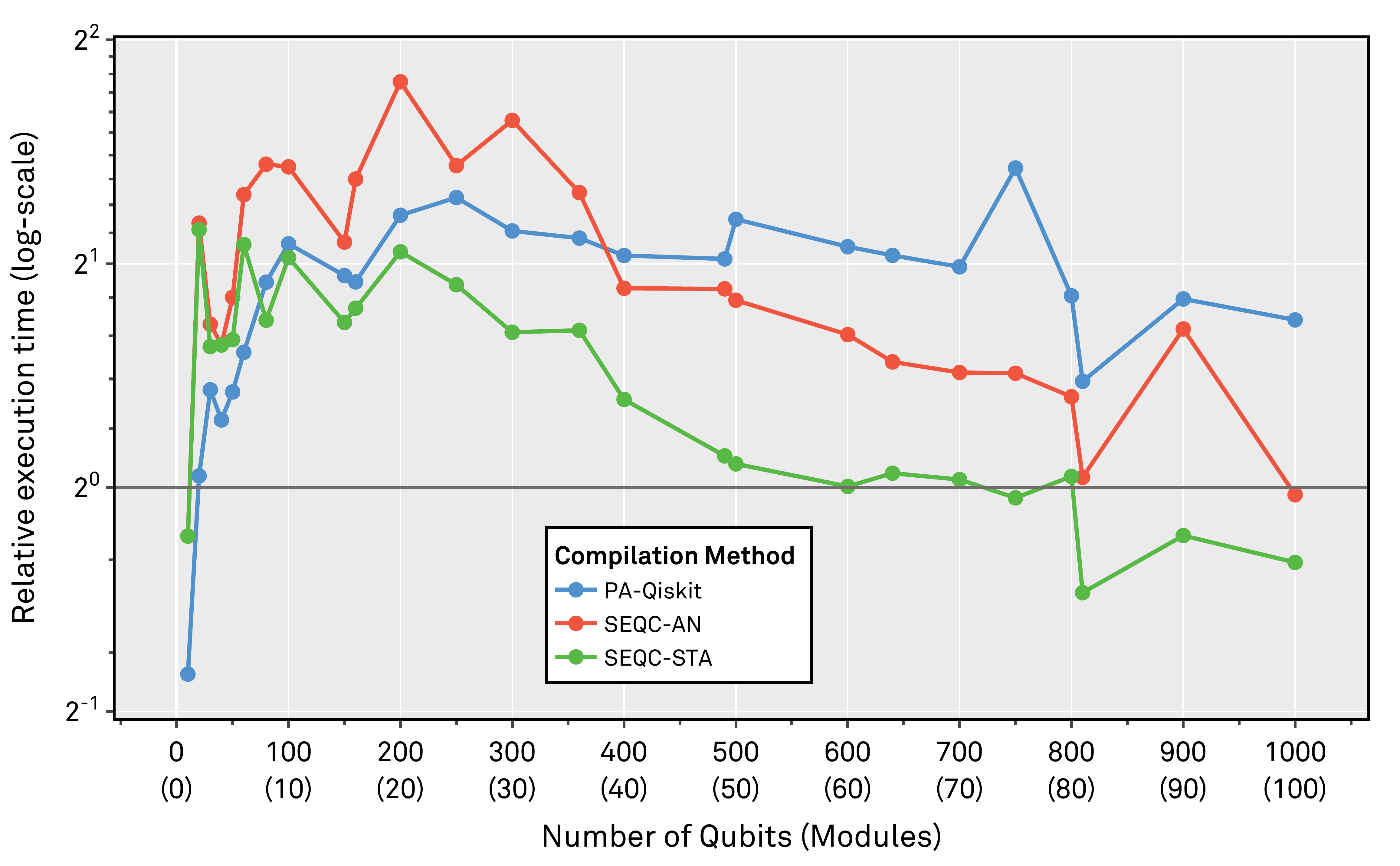}
        \subcaption{Relative per-shot execution time.}
        \label{fig:10Q_t_exec}
    \end{subfigure}
    \vspace{-0.05in}
    \caption{Total gate count, circuit depth, and per-shot execution time, 10-qubit heavy-hex backend (lower is better).}
    \label{fig:10Q_gate_count_and_depth_and_t_exec}    
\end{figure*}

\begin{figure*}[tb]
    \centering
    \begin{subfigure}[t]{0.32\linewidth}
        \centering
        \includegraphics[width=0.99\linewidth]{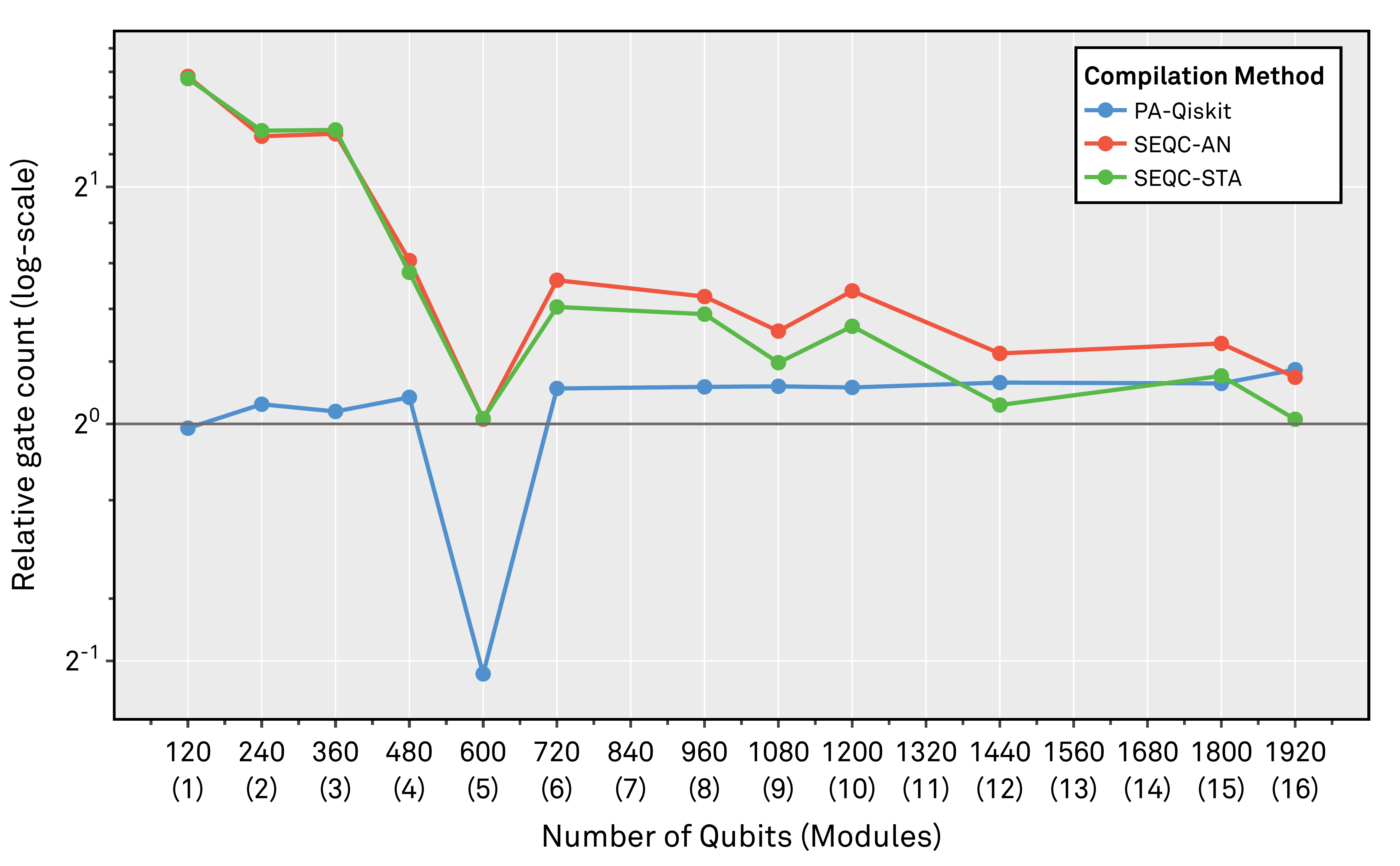}
        \subcaption{Relative total gate count.}
        \label{fig:120Q_gate_count}
    \end{subfigure}
    \begin{subfigure}[t]{0.32\linewidth}
    \centering
        \includegraphics[width=0.99\linewidth]{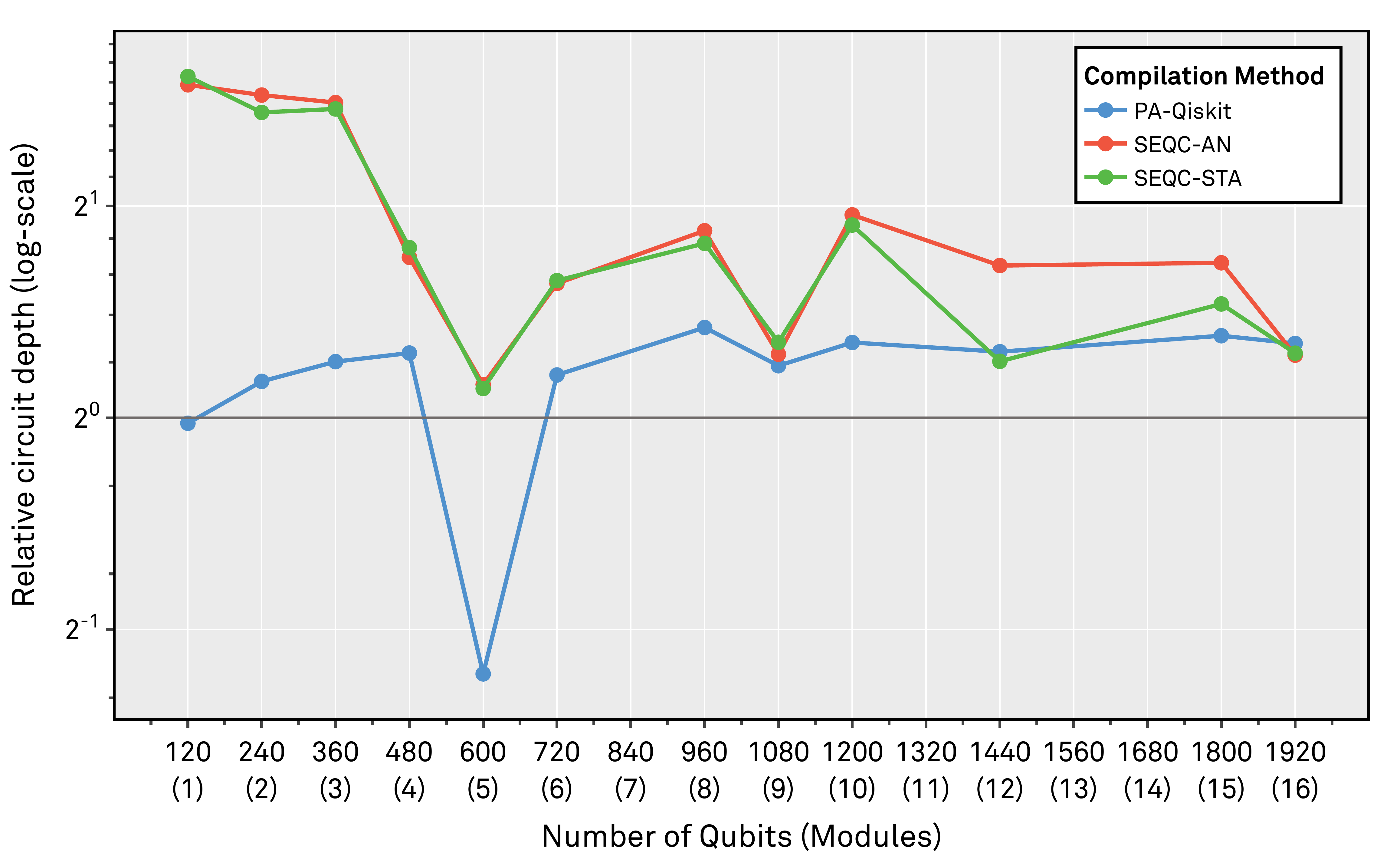}
        \subcaption{Relative circuit depth.}
        \label{fig:120Q_circuit_depth}
    \end{subfigure}
    \begin{subfigure}[t]{0.32\linewidth}
    \centering
        \includegraphics[width=0.99\linewidth]{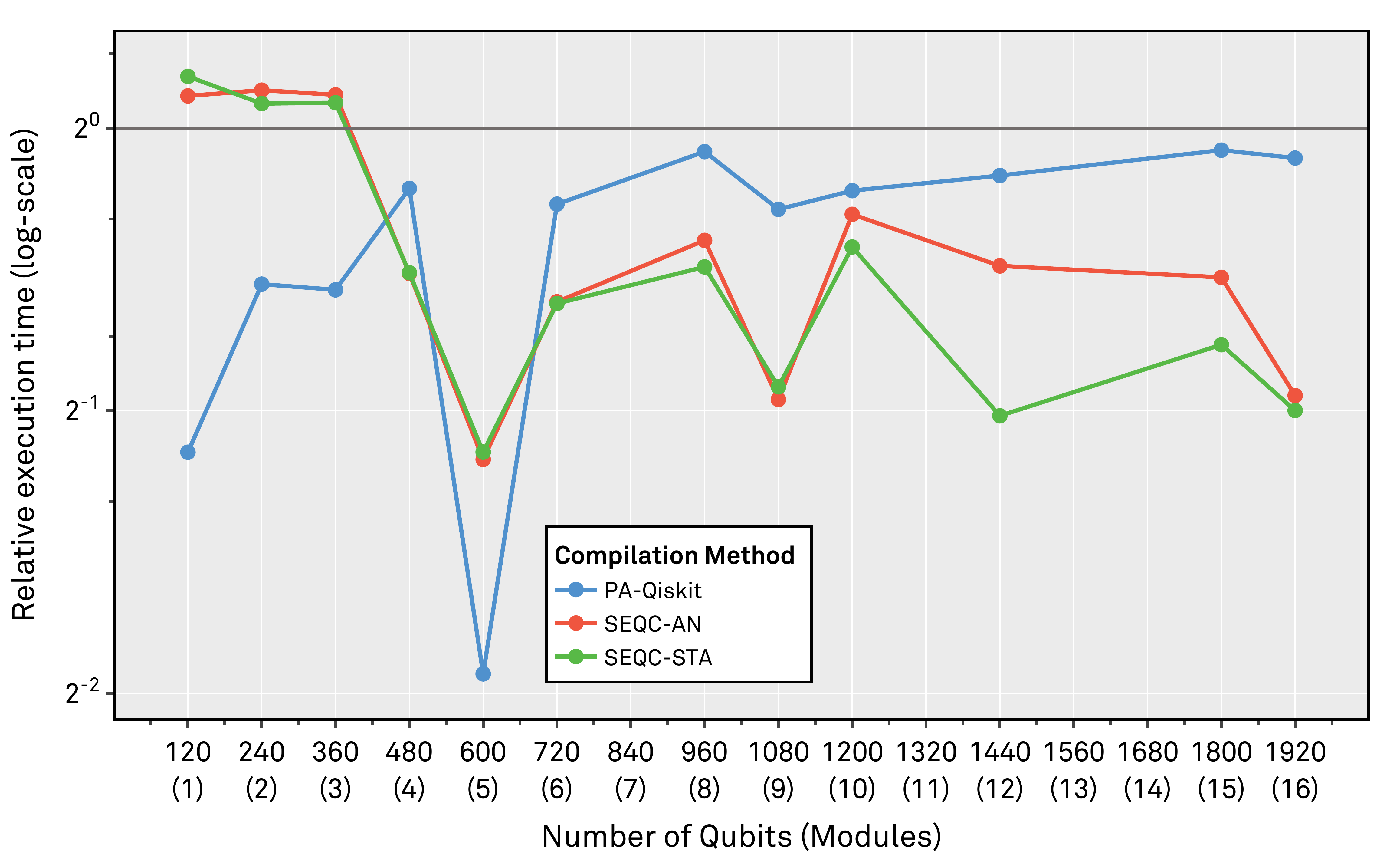}
        \subcaption{Relative per-shot execution time.}
        \label{fig:120Q_t_exec}
    \end{subfigure}
    \vspace{-0.05in}
    \caption{Total gate count, circuit depth, and per-shot execution time, 120-qubit grid backend (lower is better).}
    \label{fig:120Q_gate_count_and_depth_and_t_exec}    
\end{figure*}

\paragraph{Gate count and circuit depth} 
Gate count and circuit depth serve as static, backend-agnostic metrics for a quantum circuit's expected fidelity and execution time~\cite{mike_and_ike}. 
SABRE~\cite{zou2024lightsabrelightweightenhancedsabre,sabre_swap}, moreover, explicitly minimizes gate count by incorporating it into its cost functions.
%
We observe that SEQC typically produces results with higher gate counts (\cref{fig:10Q_gate_count,fig:120Q_gate_count}) and circuit depths (\cref{fig:10Q_circuit_depth,fig:120Q_circuit_depth}) compared to Qiskit and PA-Qiskit, as it prioritizes reducing \emph{inter-module} gates instead.
SEQC-STA, in particular, is able to demonstrate competitive results in these metrics for larger circuits, delivering comparable or slightly better results compared to PA-Qiskit at $500$ qubits (50 modules) and above. 
Additionally, SEQC-STA achieves $14.11\%$ and $8.87\%$ gate count reduction for the $810$ and $1000$ qubit ($81$ and $100$ module) compared to Qiskit, \textbf{despite Qiskit using universal inter-module links}.
Similarly, SEQC produces increasingly competitive results on the grid backend as circuit size grows; evaluating larger circuits where SEQC-STA is expected to surpass the Qiskit baselines remains future work.

\paragraph{Estimated execution time} 

The per-shot execution time of a quantum circuit is determined by the critical path of gate operations weighted by gate duration. 
It is imperative that quantum circuits execute within the decoherence time to ensure high-fidelity outcomes.
%
\cref{fig:10Q_t_exec,fig:120Q_t_exec} present estimated per-shot execution time relative to the reference, approximating inter-module SWAPs as 4$\times$ the duration of intra-module SWAPs~\cite{smith_2022_chiplets}. On the heavy-hex backend, SEQC outperforms PA-Qiskit by 24.19\% on average (up to 63.98\%), with SEQC-STA achieving execution parity with Qiskit---\textbf{despite its gate set allowances}---from 600 qubits (60 modules), and surpassing it from 810 qubits (81 modules) onwards. On the grid backend, SEQC-STA-compiled circuits execute up to 151.29\% faster than PA-Qiskit, and outperform the Qiskit baseline from 480 qubits (4 modules) by 28.04\% on average, and up to 54.82\%, with SEQC-AN trailing close behind.

\begin{figure*}[tb]
    \centering
    \begin{subfigure}[t]{0.32\linewidth}
        \centering
        \includegraphics[width=0.99\linewidth]{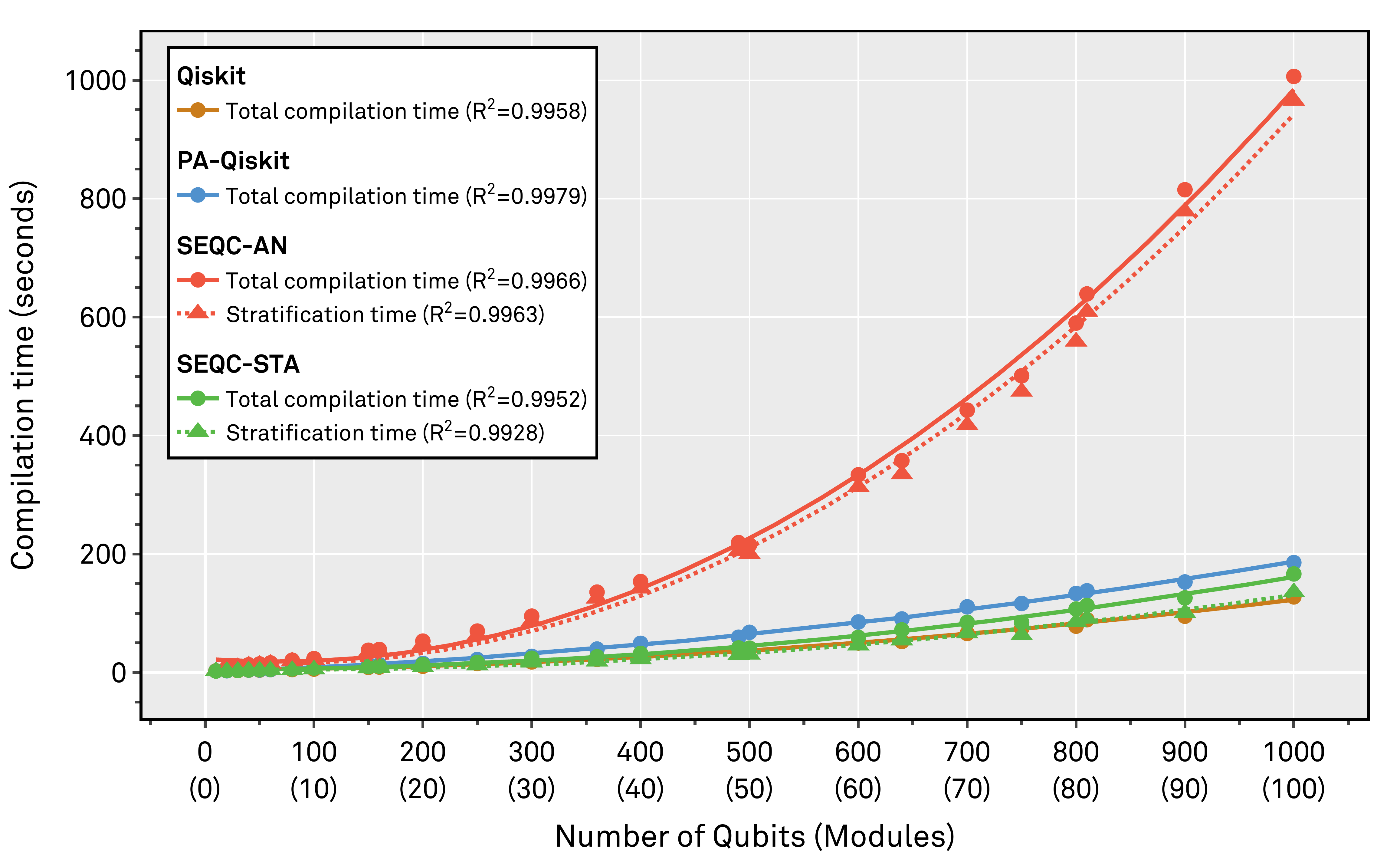}
        \subcaption{Total Compilation time.}
        \label{fig:10Q_stratification_time}
    \end{subfigure}
    \begin{subfigure}[t]{0.32\linewidth}
    \centering
        \includegraphics[width=0.99\linewidth]{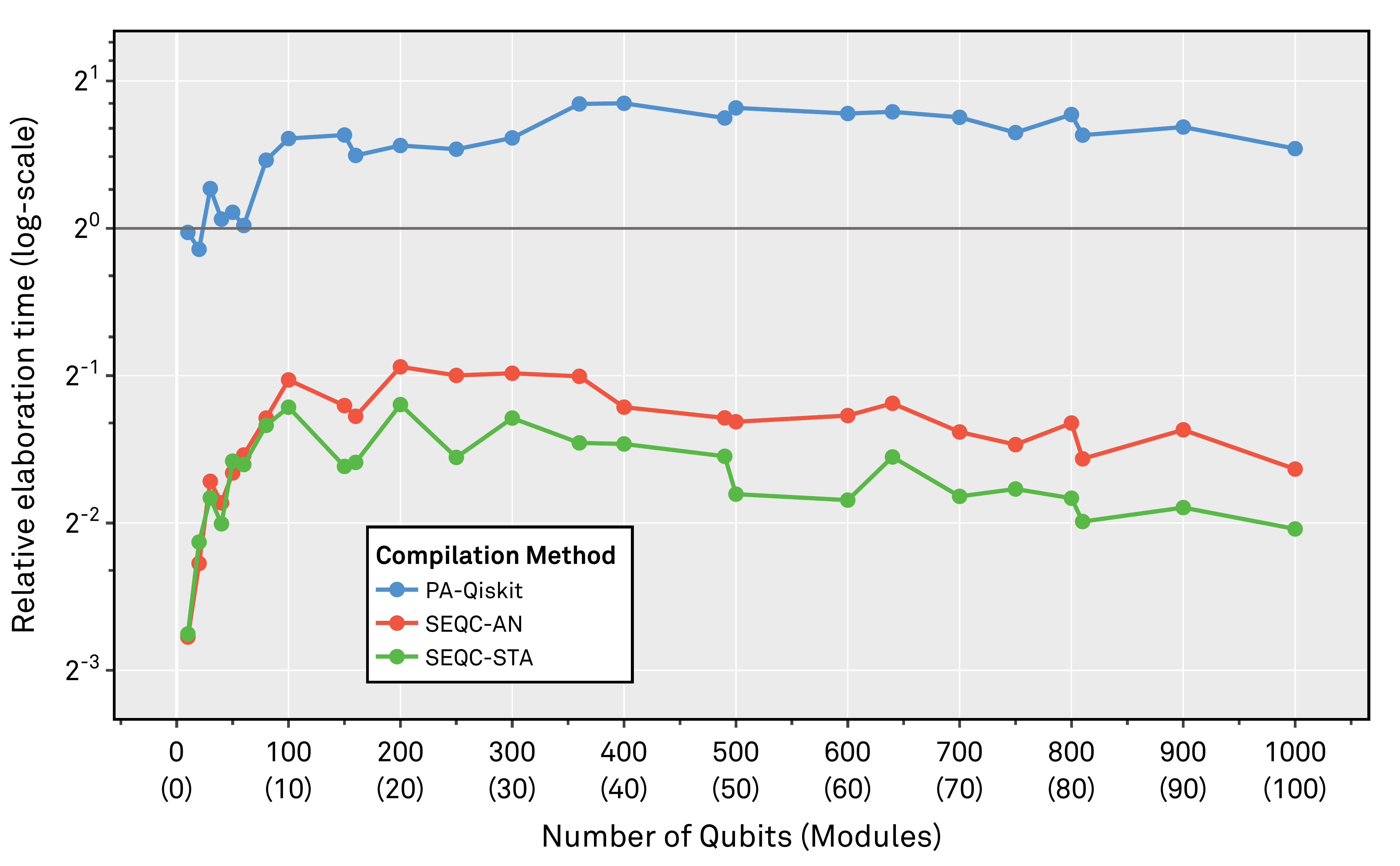}
        \subcaption{Relative recurring compilation time.}
        \label{fig:10Q_solve_time}
    \end{subfigure}
    \begin{subfigure}[t]{0.32\linewidth}
    \centering
        \includegraphics[width=0.99\linewidth]{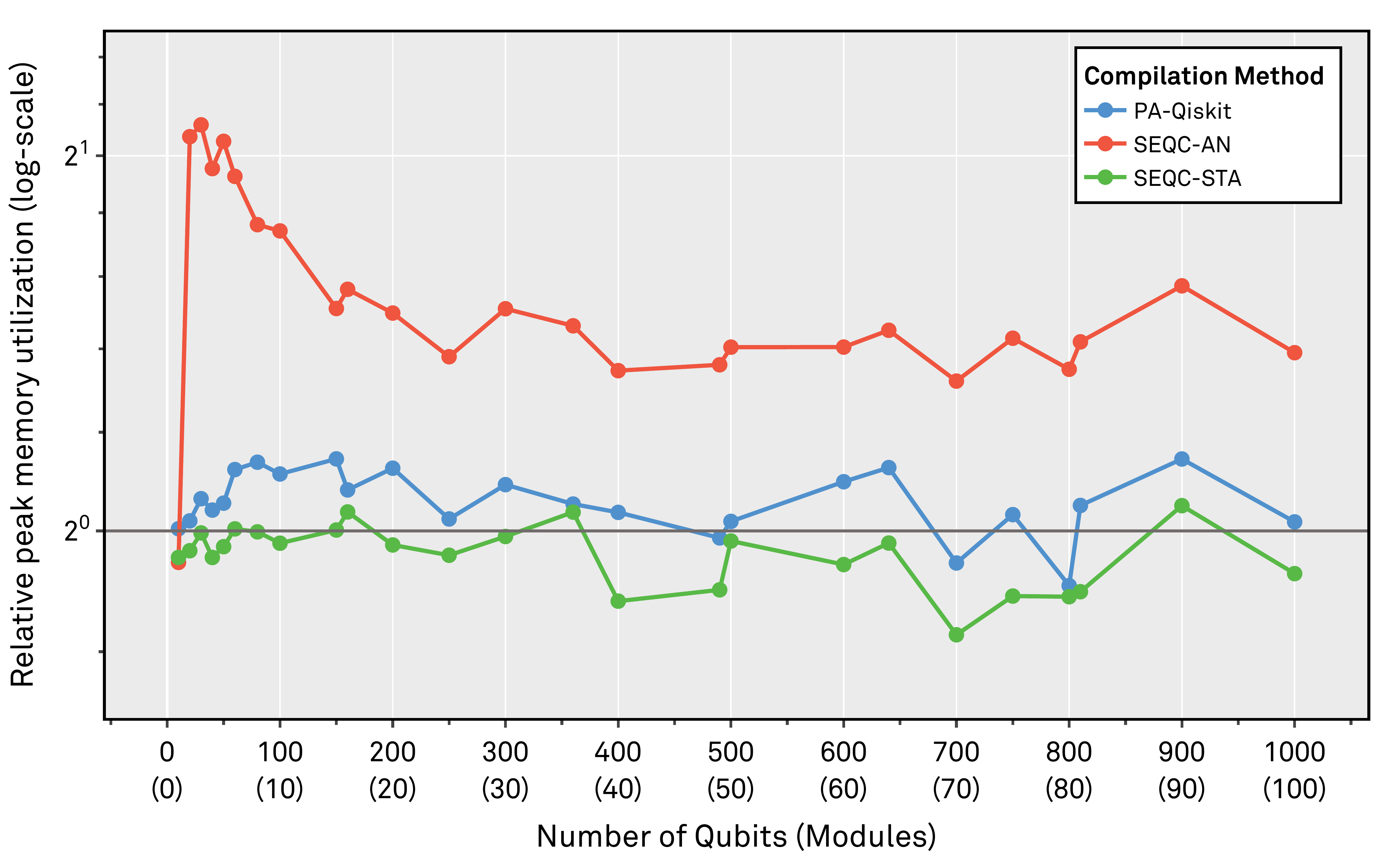}
        \subcaption{Relative peak memory utilization.}
        \label{fig:10Q_memory}
    \end{subfigure}
    \vspace{-0.05in}
    \caption{Total/recurring compilation time and peak memory utilization, 10-qubit heavy-hex backend (lower is better).}
    \label{fig:10Q_compiler_metrics}    
\end{figure*}

\begin{figure*}[tb]
    \centering
    \begin{subfigure}[t]{0.32\linewidth}
        \centering
        \includegraphics[width=0.99\linewidth]{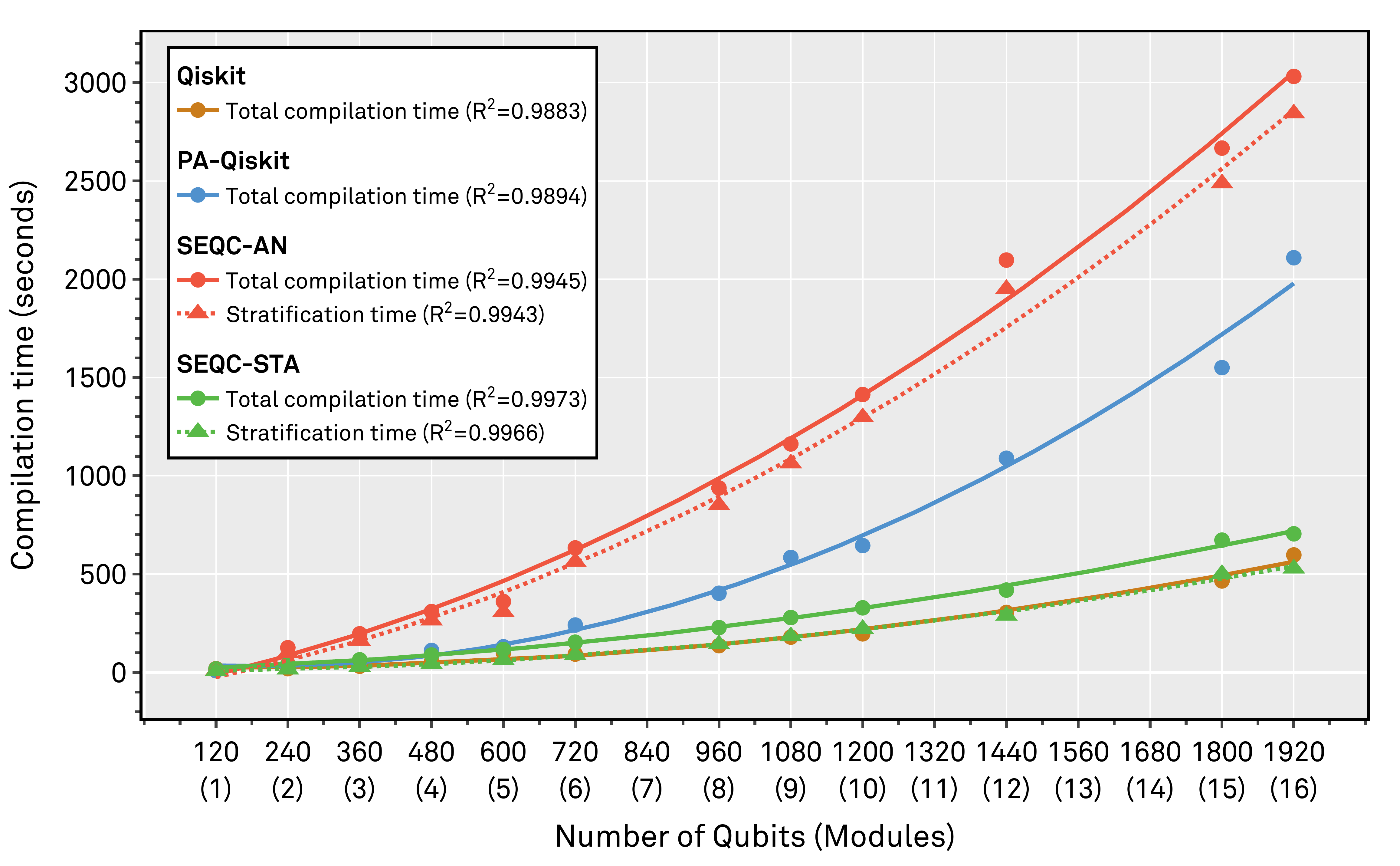}
        \subcaption{Total Compilation time.}
        \label{fig:120Q_stratification_time}
    \end{subfigure}
    \begin{subfigure}[t]{0.32\linewidth}
    \centering
        \includegraphics[width=0.99\linewidth]{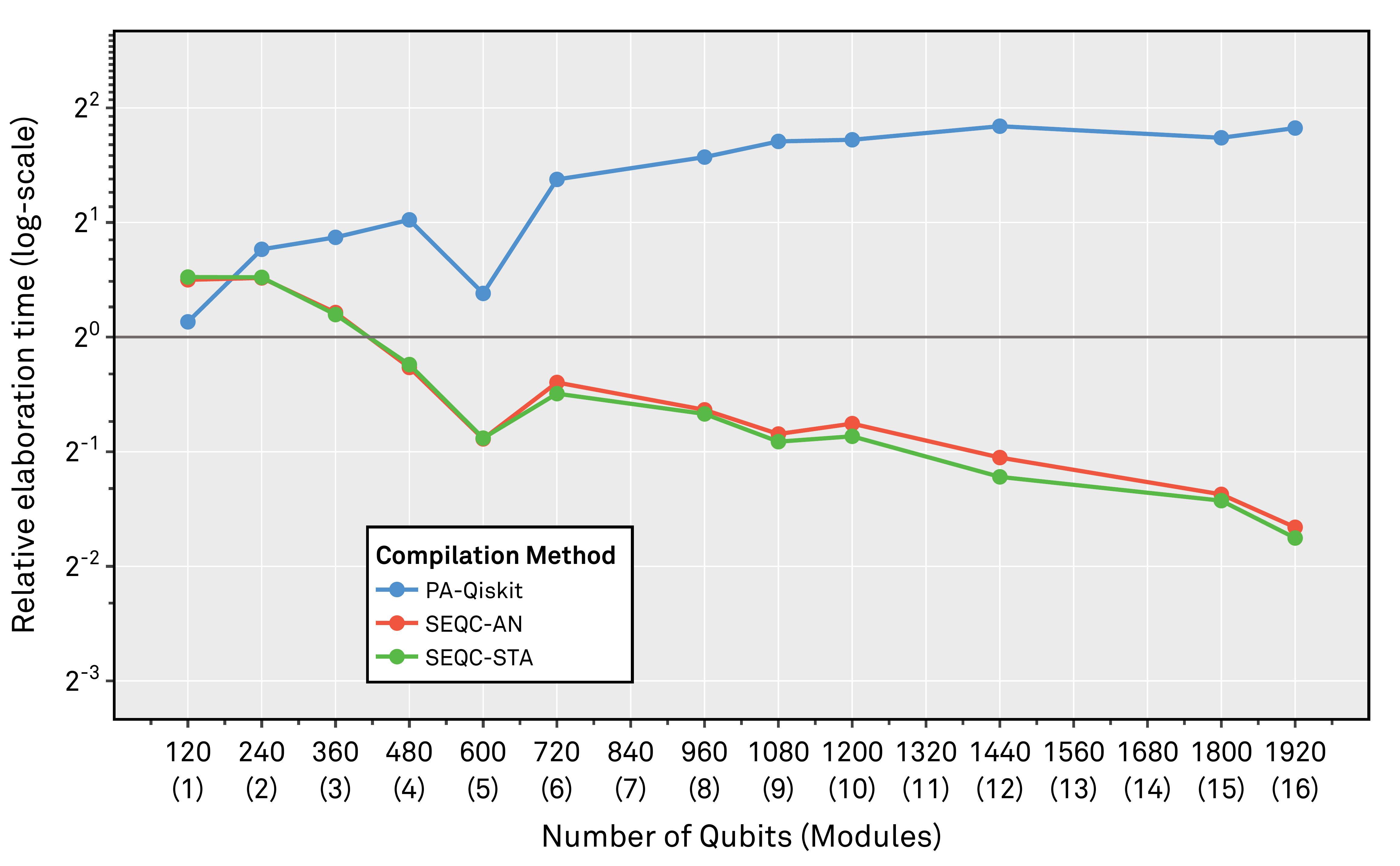}
        \subcaption{Relative recurring compilation time.}
        \label{fig:120Q_solve_time}
    \end{subfigure}
    \begin{subfigure}[t]{0.32\linewidth}
    \centering
        \includegraphics[width=0.99\linewidth]{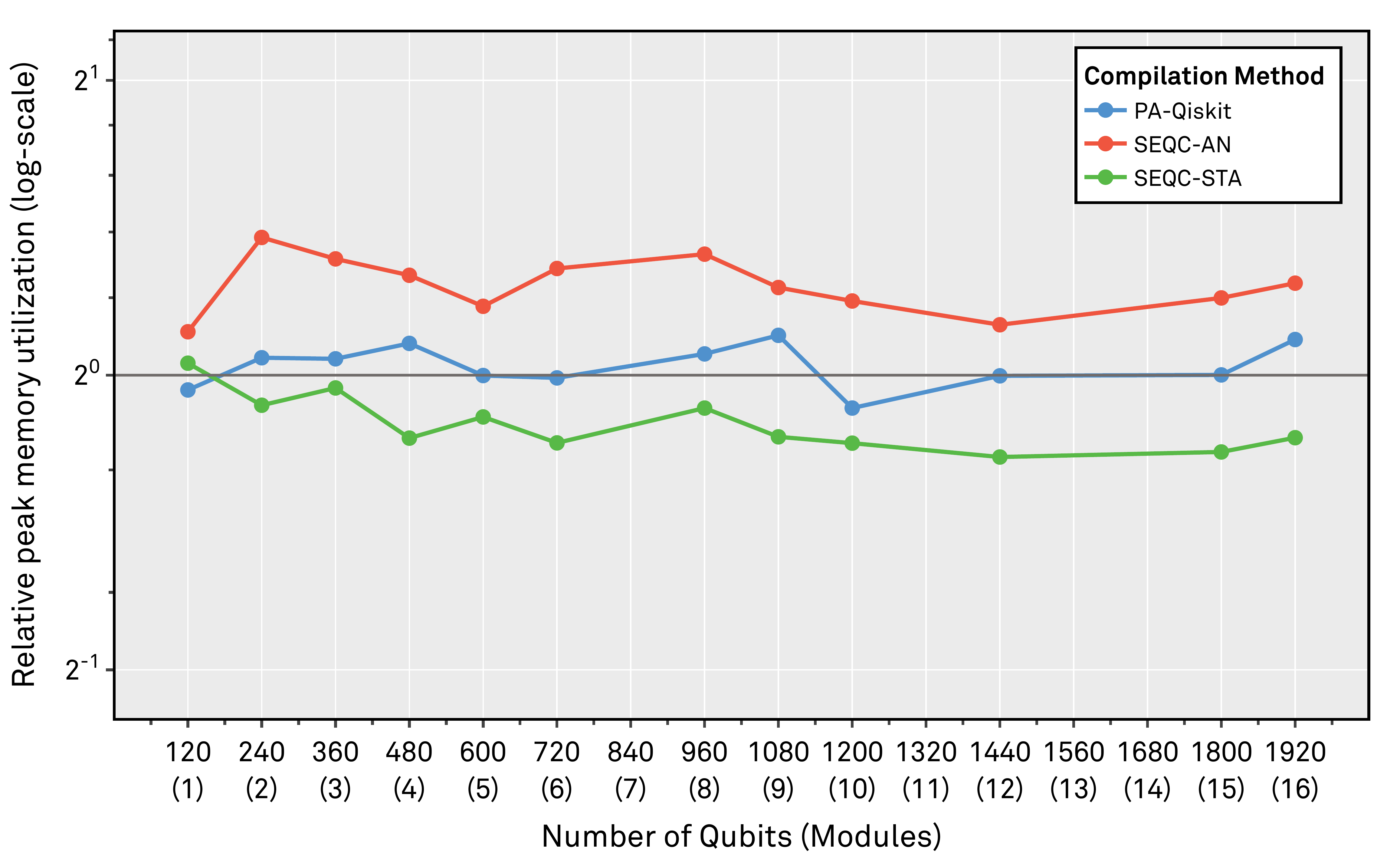}
        \subcaption{Relative peak memory utilization.}
        \label{fig:120Q_memory}
    \end{subfigure}
    \vspace{-0.05in}
    \caption{Total/recurring compilation time and peak memory utilization, 120-qubit grid backend (lower is better).}
    \label{fig:120Q_compiler_metrics}    
\end{figure*}

\paragraph{Compilation time}
While SEQC ameliorates the overhead recompilation ~\cite{Tannu_2019_ErrorVariability, Murali_2019_NoiseMapping, dasgupta2021stability} with the midway product of stratified circuits, the one-time stratification still must be scalable for practicality. 
In \cref{fig:10Q_stratification_time,fig:120Q_stratification_time}, we plot the non-recurring compilation time of our evaluated compilers. For our SEQC compilers, we plot stratification time and stratification + elaboration time. These can be compared to the total solve time for the Qiskit and PA-Qiskit compilers. We demonstrate that all compilation times obey a quadratic trajectory ($R^2 \geq 0.9883$).


To investigate SEQC's effect on recompilation time,
\cref{fig:10Q_solve_time,fig:120Q_solve_time}
show the recurring compilation time. Specifically, this refers to the elaboration time of SEQC and the full compilation time for Qiskit and PA-Qiskit. 
Overall, SEQC compilers have a large speedup over Qiskit-based compilers, owing to the use of parallelism and modular compilation strategies. 
Compared to Qiskit~/~PA-Qiskit on the heavy-hex backend, SEQC-STA achieves $3.27\times~/~4.69\times$ speedup on average, and up to $6.74\times~/~6.61\times$. Similarly on the grid backend, SEQC-STA achieves $1.34\times~/~2.52\times$ speedup on average, and up to $3.37\times~/~11.93\times$ over Qiskit~/~PA-Qiskit.


\paragraph{Memory Utilization} 
\cref{fig:10Q_memory,fig:120Q_memory} present the relative peak memory utilization of the evaluated compilers. 
Compared to Qiskit on the heavy-hex~/~grid backend, SEQC-AN exhibits $55.68\%~/~23.71\%$ higher memory usage on average, and up to $117.71\%~/~38.24\%$.
This could likely be ameliorated by reducing the number of annealing trials, at the expense of circuit quality. 
In contrast, 
SEQC-STA tends to use less memory compared to the Qiskit variants, by
$4.22\%~/~10.69\%$ on average, and up to $17.47\%~/~17.51\%$, on the heavy-hex~/~grid backends, respectively.
The memory conservation likely comes from stratification limiting the subcircuit sizes at the elaboration stage, presuming that the memory requirements scales super-linearly with the circuit size. This points to the scalability benefits of the SEQC framework in general.





\section{Discussion} 
\label{sec:discussion}

These results expose a fundamental flaw in using gate count and circuit depth as heuristics: both treat all gates equally, ignoring that two-qubit gates are significantly noisier than one-qubit gates, and inter-module gates noisier still~\cite{smith_2022_chiplets}. SEQC addresses this implicitly by prioritizing inter-module gate minimization over total gate count. Stratification and elaboration together confer noise-awareness, yielding better optimization of circuit fidelity and execution time.

Overall, we observe that SEQC compiles faster and yields circuits with more desirable characteristics for modular quantum architectures.
These improvements widen as the circuit's number of qubits increases, especially for the most important metrics of fidelity, compiled circuit execution time, and number of inter-module gates. 
These trends hold promise in allowing for faster and better compilation for future large-scale, hybrid modular quantum devices than we can achieve with today's frameworks.

It is also important to note that SEQC is equally applicable to NISQ as well as future quantum systems. The stratification stage can decompose a large circuit into subcircuits that will be mapped to QPUs at the virtual level, and then the elaboration stage can compile for each QPU together with an error correction method for that QPU. 
Applicability to NISQ systems is still important, as the transition to Fault Tolerant Quantum Computing (FTQC) will be gradual. Early fault-tolerant demonstrations will still heavily rely on NISQ-era methods, as only few qubits can be protected.
Moreover, as recent results demonstrate, practical scientific and industrial value exist for NISQ and early fault tolerant systems, without having access to full error-correction~\cite{piccinelli_chemistry, Kim2023preFTUtility, sciadv_utility_2025, Abanin2025ConstructiveInterference}. 
Thus, it remains critical to design architectures and compilers that serve NISQ hardware, both to extract near-term value, and to pave the engineering path toward full FTQC.

\section{Related Work} \label{sec:related_work}


\textbf{Distributed quantum computing}
(DQC)~\cite{autocomm, 10228915,
10.1145/3730585,
10821229,
10313645,
10.1145/3579367} 
requires generating and distributing EPR pairs over long-distance optical links.
In contrast, modular architectures have less restrictive distance constraints and often support other hardware technologies, e.g., microwave links~\cite{Field_2024_RigettiModular}.
As a result, SEQC can model inter-module allocation topologically as SWAP chains~\cite{LaRacuente2025modelingshortrange}, similar to monolithic quantum compilation. 
%

\textbf{Trapped-ion device compilation.}
Compilation for modular systems shares many similar constraints with compilation for trapped-ion devices. 
However, some key differences make it difficult to compare SEQC with trapped-ion compilers on equal footing. 
Notably, trapped-ion compilers~\cite{10.1145/3526241.3530366,Kreppel2023quantumcircuit} can employ simpler routing strategies, or forgo them entirely, compared to generic modular compilers, due to their less restrictive topological constraints in applying two-qubit operations.

\textbf{Complementary Approaches.}
MECH~\cite{mech} entangles ancillary qubits and distributes them throughout a module to build fast network-on-chip routing channels (a ``\textit{communication highway}''). The highway, however, does not generalize to non-universal inter-module links. 
SEQC is complementary to MECH and could be extended with its methodology to gain the benefits of both.
%
Lin \textit{et al.}~\cite{let_every_qubit} use calibration data to determine potentially nonstandard two-qubit basis gates more compatible with their respective physical states and improve fidelity.
%
Bandic \textit{et al.}\cite{bandic2024profiling} leverage quantum circuit structure and mapping efficiency to optimize the architecture of both monolithic and modular quantum processors.
%
Escofet \textit{et al.}~\cite{escofet2024revisiting} develop a characterization technique to assess the optimality of qubit mapping.
%
These techniques are orthogonal to SEQC, and could work synergistically with it.

\textbf{Circuit Cutting.} 
CutQC~\cite{tang2021cutqc} partitions large quantum circuits into smaller subcircuits that fit on a device, executes them individually, and then classically stitches the results. 
Similarly, Piveteau and Sutter~\cite{circuit_knitting} uses quasiprobability simulation to model inter-module link behavior between partitioned subcircuits.
In contrast, SEQC does not perform any classical stitching or simulation. We originally formulated a constraint optimization problem, similar to CutQC, but numerical solvers were impractically slow for circuits over 20 qubits, corroborating the experience of Baker \textit{et al.}~\cite{baker2020}. 
%
Bandic \textit{et al.}~\cite{bandic2023mapping} map circuit partitioning to \textsc{Min-Cut} and use the QUBO model with quantum solvers to generate a set of cuts intended to minimize SWAPs between partitions. 
Instead, SEQC takes a ``best-effort'' approach that works well in practice, and does not require acceleration through quantum means.

\textbf{Related Qubit Allocation Works.} 
Liu \textit{et al.}~\cite{liu_2023_tackling} partition a circuit into blocks with equal number of qubits, perform permutation-aware synthesis on each block in parallel across all possible qubit topologies, and map the blocks to the target device with an augmented SABRE algorithm. SEQC instead settles SWAP permutations between modules  prior to any per-module compilation, and, as a result, requires only solving one qubit layout per block rather than all permutations.
%
Hungarian Qubit Assignment (HQA)~\cite{escofet2023hungarian} and Pastor \textit{et al.}~\cite{pastor2024circuit} also tackle modular compilation, 
but assume that all modules are fully connected, which is both impractical for large-scale systems and misses optimization opportunities. SEQC, in contrast, can work with arbitrary module topologies.
Ovide \textit{et al.}~\cite{ovide2023mapping} leverage both quantum and classical methods for inter-module communication.
While it advocates for a two-level approach to quantum circuit mapping, it leaves the actual problem of qubit placement and routing unsolved, which SEQC addresses.

\section{Conclusion}
The increasing complexity and scale of modular quantum devices challenge quantum compilers.
We propose SEQC, which implements several novel module- and parallelization-aware methods for qubit placement, qubit routing, and circuit optimization. 
SEQC achieves on average 9.3--32.3\% higher circuit fidelity (up to 49.99--63.36\%), depending on the module size and topology, attaining an average 1.34--3.27$\times$ faster compilation (up to 3.37--6.74$\times$) compared to a module-unaware Qiskit baseline. These results are demonstrated on homogeneous modules to isolate the impact of our techniques and enable comparison with existing compilers; extending SEQC to hybrid QPU systems is natural with module-specific elaboration stages, while leaving other core techniques intact.



%% file: refs.bib
@IEEEtranBSTCTL{IEEEexample:BSTcontrol,
  CTLuse_forced_etal       = "yes",
  CTLmax_names_forced_etal = "6",
  CTLnames_show_etal       = "1"
}

@inproceedings{smith_2022_chiplets,
  author={Smith, Kaitlin N. and Ravi, Gokul Subramanian and Baker, Jonathan M. and Chong, Frederic T.},
  booktitle={2022 55th IEEE/ACM International Symposium on Microarchitecture (MICRO)}, 
  title={Scaling Superconducting Quantum Computers with Chiplet Architectures}, 
  year={2022},
  volume={},
  number={},
  pages={1092-1109},
  doi={10.1109/MICRO56248.2022.00078}
}

@inproceedings{supermarq,
author = {T. Tomesh and P. Gokhale and V. Omole and G. Ravi and K. N. Smith and J. Viszlai and X. Wu and N. Hardavellas and M. R. Martonosi and F. T. Chong},
booktitle = {2022 IEEE International Symposium on High-Performance Computer Architecture (HPCA)},
title = {SupermarQ: A Scalable Quantum Benchmark Suite},
year = {2022},
volume = {},
issn = {},
pages = {587-603},
doi = {10.1109/HPCA53966.2022.00050},
url = {https://doi.ieeecomputersociety.org/10.1109/HPCA53966.2022.00050},
publisher = {IEEE Computer Society},
address = {Los Alamitos, CA, USA},
month = {apr}
}

@misc{zou2024lightsabrelightweightenhancedsabre,
      title={Light{SABRE}: A Lightweight and Enhanced {SABRE} Algorithm}, 
      author={Henry Zou and Matthew Treinish and Kevin Hartman and Alexander Ivrii and Jake Lishman},
      year={2024},
      eprint={2409.08368},
      archivePrefix={arXiv},
      primaryClass={quant-ph},
      url={https://arxiv.org/abs/2409.08368}, 
}

@inproceedings{baker2020,
author = {Baker, Jonathan M. and Duckering, Casey and Hoover, Alexander and Chong, Frederic T.},
title = {Time-sliced quantum circuit partitioning for modular architectures},
year = {2020},
isbn = {9781450379564},
publisher = {Association for Computing Machinery},
address = {New York, NY, USA},
url = {https://doi.org/10.1145/3387902.3392617},
doi = {10.1145/3387902.3392617},
booktitle = {Proceedings of the 17th ACM International Conference on Computing Frontiers},
pages = {98–107},
numpages = {10},
location = {Catania, Sicily, Italy},
series = {CF '20}
}

@INPROCEEDINGS{autocomm,
  author={Wu, Anbang and Zhang, Hezi and Li, Gushu and Shabani, Alireza and Xie, Yuan and Ding, Yufei},
  booktitle={2022 55th IEEE/ACM International Symposium on Microarchitecture (MICRO)}, 
  title={Auto{C}omm: A Framework for Enabling Efficient Communication in Distributed Quantum Programs}, 
  year={2022},
  volume={},
  number={},
  pages={1027-1041},
  doi={10.1109/MICRO56248.2022.00074}}

@article{heavy_hex,
  title = {Topological and Subsystem Codes on Low-Degree Graphs with Flag Qubits},
  author = {Chamberland, Christopher and Zhu, Guanyu and Yoder, Theodore J. and Hertzberg, Jared B. and Cross, Andrew W.},
  journal = {Phys. Rev. X},
  volume = {10},
  issue = {1},
  pages = {011022},
  numpages = {19},
  year = {2020},
  month = {Jan},
  publisher = {American Physical Society},
  doi = {10.1103/PhysRevX.10.011022},
  url = {https://link.aps.org/doi/10.1103/PhysRevX.10.011022}
}

@inproceedings{qfactor,
  author={Kukliansky, Alon and Younis, Ed and Cincio, Lukasz and Iancu, Costin},
  booktitle={2023 IEEE International Conference on Quantum Computing and Engineering (QCE)}, 
  title={{QF}actor: A Domain-Specific Optimizer for Quantum Circuit Instantiation}, 
  year={2023},
  volume={01},
  number={},
  pages={814-824},
  doi={10.1109/QCE57702.2023.00096}
}

@article{esp,
author={Qi, Fang and Smith, Kaitlin N. and LeCompte, Travis and Tzeng, Nian-feng and Yuan, Xu and Chong, Frederic T. and Peng, Lu},
journal={ IEEE Transactions on Quantum Engineering },
title={Quantum Vulnerability Analysis to Guide Robust Quantum Computing System Design},
year={2024},
volume={5},
number={01},
ISSN={2689-1808},
pages={1-11},
doi={10.1109/TQE.2023.3343625},
url = {https://doi.ieeecomputersociety.org/10.1109/TQE.2023.3343625},
publisher={IEEE Computer Society},
address={Los Alamitos, CA, USA},
month={01}
}

@misc{bqskit,
title = {Berkeley Quantum Synthesis Toolkit ({BQSKit}) v1},
author = {Younis, Ed and Iancu, Costin C and Lavrijsen, Wim and Davis, Marc and Smith, Ethan and USDOE},
abstractNote = {The Berkeley Quantum Synthesis Toolkit (BQSKit) is an optimizing quantum compiler and research vehicle that combines ideas from several projects at LBNL into one easily accessible and quickly extensible software package. The ideas in the QFAST, QSearch, LEAP, and QFactor software tools (all licensed through ipo.lbl.gov) all build upon one another. By combining these into one package, we create symbiotic interactions between the tools. This means better results, better throughput, less to maintain, and greater surface area to the public. Additionally, the BQSKit tool will create a research platform for future work here at LBNL.},
doi = {10.11578/dc.20210603.2},
url = {https://www.osti.gov/biblio/1785933}, 
year = {2021},
month = {4},
}

@inproceedings{qgo,
  author={Wu, Xin-Chuan and Davis, Marc Grau and Chong, Frederic T. and Iancu, Costin},
  booktitle={2021 International Conference on Rebooting Computing (ICRC)}, 
  title={Reoptimization of Quantum Circuits via Hierarchical Synthesis}, 
  year={2021},
  volume={},
  number={},
  pages={35-46},
  doi={10.1109/ICRC53822.2021.00016}}

@article{optimization1,
  author={Maslov, Dmitri and Dueck, Gerhard W. and Miller, D. Michael and Negrevergne, Camille},
  journal={IEEE Transactions on Computer-Aided Design of Integrated Circuits and Systems}, 
  title={Quantum Circuit Simplification and Level Compaction}, 
  year={2008},
  volume={27},
  number={3},
  pages={436-444},
  doi={10.1109/TCAD.2007.911334}}

@article{paler2019really,
  title={Really small shoe boxes: On realistic quantum resource estimation},
  author={Paler, Alexandru and Herr, Daniel and Devitt, Simon J},
  journal={Computer},
  volume={52},
  number={6},
  pages={27--37},
  year={2019},
  publisher={IEEE}
}

@article{circuit_knitting,
  author={Piveteau, Christophe and Sutter, David},
  journal={IEEE Transactions on Information Theory}, 
  title={Circuit Knitting With Classical Communication}, 
  year={2024},
  volume={70},
  number={4},
  pages={2734-2745},
  doi={10.1109/TIT.2023.3310797}
}

@inproceedings{sabre_swap,
author = {Li, Gushu and Ding, Yufei and Xie, Yuan},
title = {Tackling the Qubit Mapping Problem for {NISQ}-Era Quantum Devices},
year = {2019},
isbn = {9781450362405},
publisher = {Association for Computing Machinery},
address = {New York, NY, USA},
url = {https://doi.org/10.1145/3297858.3304023},
doi = {10.1145/3297858.3304023},
booktitle = {Proceedings of the Twenty-Fourth International Conference on Architectural Support for Programming Languages and Operating Systems},
pages = {1001–1014},
numpages = {14},
location = {Providence, RI, USA},
series = {ASPLOS '19}
}

@inproceedings{tang2021cutqc,
  title={{CutQC}: using small quantum computers for large quantum circuit evaluations},
  author={Tang, Wei and Tomesh, Teague and Suchara, Martin and Larson, Jeffrey and Martonosi, Margaret},
  booktitle={Proceedings of the 26th ACM International conference on architectural support for programming languages and operating systems},
  pages={473--486},
  year={2021}
}

@article{backend_specs,
author={Acharya, Rajeev
and Aleiner, Igor
and Allen, Richard
and Andersen, Trond I.
and Ansmann, Markus
and Arute, Frank
and Arya, Kunal
and Asfaw, Abraham
and Atalaya, Juan
and Babbush, Ryan
and Bacon, Dave
and Bardin, Joseph C.
and Basso, Joao
and Bengtsson, Andreas
and Boixo, Sergio
and Bortoli, Gina
and Bourassa, Alexandre
and Bovaird, Jenna
and Brill, Leon
and Broughton, Michael
and Buckley, Bob B.
and Buell, David A.
and Burger, Tim
and Burkett, Brian
and Bushnell, Nicholas
and Chen, Yu
and Chen, Zijun
and Chiaro, Ben
and Cogan, Josh
and Collins, Roberto
and Conner, Paul
and Courtney, William
and Crook, Alexander L.
and Curtin, Ben
and Debroy, Dripto M.
and Del Toro Barba, Alexander
and Demura, Sean
and Dunsworth, Andrew
and Eppens, Daniel
and Erickson, Catherine
and Faoro, Lara
and Farhi, Edward
and Fatemi, Reza
and Flores Burgos, Leslie
and Forati, Ebrahim
and Fowler, Austin G.
and Foxen, Brooks
and Giang, William
and Gidney, Craig
and Gilboa, Dar
and Giustina, Marissa
and Grajales Dau, Alejandro
and Gross, Jonathan A.
and Habegger, Steve
and Hamilton, Michael C.
and Harrigan, Matthew P.
and Harrington, Sean D.
and Higgott, Oscar
and Hilton, Jeremy
and Hoffmann, Markus
and Hong, Sabrina
and Huang, Trent
and Huff, Ashley
and Huggins, William J.
and Ioffe, Lev B.
and Isakov, Sergei V.
and Iveland, Justin
and Jeffrey, Evan
and Jiang, Zhang
and Jones, Cody
and Juhas, Pavol
and Kafri, Dvir
and Kechedzhi, Kostyantyn
and Kelly, Julian
and Khattar, Tanuj
and Khezri, Mostafa
and Kieferov{\'a}, M{\'a}ria
and Kim, Seon
and Kitaev, Alexei
and Klimov, Paul V.
and Klots, Andrey R.
and Korotkov, Alexander N.
and Kostritsa, Fedor
and Kreikebaum, John Mark
and Landhuis, David
and Laptev, Pavel
and Lau, Kim-Ming
and Laws, Lily
and Lee, Joonho
and Lee, Kenny
and Lester, Brian J.
and Lill, Alexander
and Liu, Wayne
and Locharla, Aditya
and Lucero, Erik
and Malone, Fionn D.
and Marshall, Jeffrey
and Martin, Orion
and McClean, Jarrod R.
and McCourt, Trevor
and McEwen, Matt
and Megrant, Anthony
and Meurer Costa, Bernardo
and Mi, Xiao
and Miao, Kevin C.
and Mohseni, Masoud
and Montazeri, Shirin
and Morvan, Alexis
and Mount, Emily
and Mruczkiewicz, Wojciech
and Naaman, Ofer
and Neeley, Matthew
and Neill, Charles
and Nersisyan, Ani
and Neven, Hartmut
and Newman, Michael
and Ng, Jiun How
and Nguyen, Anthony
and Nguyen, Murray
and Niu, Murphy Yuezhen
and O'Brien, Thomas E.
and Opremcak, Alex
and Platt, John
and Petukhov, Andre
and Potter, Rebecca
and Pryadko, Leonid P.
and Quintana, Chris
and Roushan, Pedram
and Rubin, Nicholas C.
and Saei, Negar
and Sank, Daniel
and Sankaragomathi, Kannan
and Satzinger, Kevin J.
and Schurkus, Henry F.
and Schuster, Christopher
and Shearn, Michael J.
and Shorter, Aaron
and Shvarts, Vladimir
and Skruzny, Jindra
and Smelyanskiy, Vadim
and Smith, W. Clarke
and Sterling, George
and Strain, Doug
and Szalay, Marco
and Torres, Alfredo
and Vidal, Guifre
and Villalonga, Benjamin
and Vollgraff Heidweiller, Catherine
and White, Theodore
and Xing, Cheng
and Yao, Z. Jamie
and Yeh, Ping
and Yoo, Juhwan
and Young, Grayson
and Zalcman, Adam
and Zhang, Yaxing
and Zhu, Ningfeng
and AI, Google Quantum},
title={Suppressing quantum errors by scaling a surface code logical qubit},
journal={Nature},
year={2023},
month={Feb},
day={01},
volume={614},
number={7949},
pages={676-681},
issn={1476-4687},
doi={10.1038/s41586-022-05434-1},
url={https://doi.org/10.1038/s41586-022-05434-1}
}

@inproceedings{qubit_allocation,
author = {Siraichi, Marcos Yukio and Santos, Vin\'{\i}cius Fernandes dos and Collange, Caroline and Pereira, Fernando Magno Quintao},
title = {Qubit allocation},
year = {2018},
isbn = {9781450356176},
publisher = {Association for Computing Machinery},
address = {New York, NY, USA},
url = {https://doi.org/10.1145/3168822},
doi = {10.1145/3168822},
booktitle = {Proceedings of the 2018 International Symposium on Code Generation and Optimization},
pages = {113–125},
numpages = {13},
location = {Vienna, Austria},
series = {CGO 2018}
}

@inproceedings{qubit_routing,
author="Ito, Takehiro
and Kakimura, Naonori
and Kamiyama, Naoyuki
and Kobayashi, Yusuke
and Okamoto, Yoshio",
editor="Morin, Pat
and Suri, Subhash",
title="Algorithmic Theory of Qubit Routing",
booktitle="Algorithms and Data Structures",
year="2023",
publisher="Springer Nature Switzerland",
address="Cham",
pages="533--546",
isbn="978-3-031-38906-1"
}

@inproceedings{spec2017,
author = {Bucek, James and Lange, Klaus-Dieter and v. Kistowski, J\'{o}akim},
title = {{SPEC CPU}2017: Next-Generation Compute Benchmark},
year = {2018},
isbn = {9781450356299},
publisher = {Association for Computing Machinery},
address = {New York, NY, USA},
url = {https://doi.org/10.1145/3185768.3185771},
doi = {10.1145/3185768.3185771},
booktitle = {Companion of the 2018 ACM/SPEC International Conference on Performance Engineering},
pages = {41–42},
numpages = {2},
location = {Berlin, Germany},
series = {ICPE '18}
}

@misc{mech,
  title={{MECH}: Multi-Entry Communication Highway for Superconducting Quantum Chiplets}, 
  author={Hezi Zhang and Keyi Yin and Anbang Wu and Hassan Shapourian and Alireza Shabani and Yufei Ding},
  year={2024},
  eprint={2305.05149},
  archivePrefix={arXiv},
  primaryClass={quant-ph}
}

@misc{escofet2024revisiting,
      title={Revisiting the Mapping of Quantum Circuits: Entering the Multi-Core Era}, 
      author={Pau Escofet and Anabel Ovide and Medina Bandic and Luise Prielinger and Hans van Someren and Sebastian Feld and Eduard Alarcón and Sergi Abadal and Carmen G. Almudéver},
      year={2024},
      eprint={2403.17205},
      archivePrefix={arXiv},
      primaryClass={quant-ph}
}

@inproceedings{bandic2023mapping,
  title={Mapping quantum circuits to modular architectures with {QUBO}},
  author={Bandic, Medina and Prielinger, Luise and N{\"u}{\ss}lein, Jonas and Ovide, Anabel and Rodrigo, Santiago and Abadal, Sergi and van Someren, Hans and Vardoyan, Gayane and Alarcon, Eduard and Almudever, Carmen G and others},
  booktitle={2023 IEEE International Conference on Quantum Computing and Engineering (QCE)},
  volume={1},
  pages={790--801},
  year={2023},
  organization={IEEE}
}

@inproceedings{ovide2023mapping,
  title={Mapping quantum algorithms to multi-core quantum computing architectures},
  author={Ovide, Anabel and Rodrigo, Santiago and Bandic, Medina and Van Someren, Hans and Feld, Sebastian and Abadal, Sergi and Alarcon, Eduard and Almudever, Carmen G},
  booktitle={2023 IEEE International Symposium on Circuits and Systems (ISCAS)},
  pages={1--5},
  year={2023},
  organization={IEEE}
}

@article{escofet2023hungarian,
  title={Hungarian qubit assignment for optimized mapping of quantum circuits on multi-core architectures},
  author={Escofet, Pau and Ovide, Anabel and Almudever, Carmen G and Alarc{\'o}n, Eduard and Abadal, Sergi},
  journal={IEEE Computer Architecture Letters},
  year={2023},
  publisher={IEEE}
}

@article{pastor2024circuit,
  title={Circuit Partitioning for Multi-Core Quantum Architectures with Deep Reinforcement Learning},
  author={Pastor, Arnau and Escofet, Pau and Rached, Sahar Ben and Alarc{\'o}n, Eduard and Barlet-Ros, Pere and Abadal, Sergi},
  journal={arXiv preprint arXiv:2401.17976},
  year={2024}
}

@article{Solovay-Kitaev,
    author = {Dawson, Christopher M. and Nielsen, Michael A.},
    title = {The {S}olovay-{K}itaev algorithm},
    year = {2006},
    issue_date = {January 2006},
    publisher = {Rinton Press, Incorporated},
    address = {Paramus, NJ},
    volume = {6},
    number = {1},
    issn = {1533-7146},
    journal = {Quantum Info. Comput.},
    month = {jan},
    pages = {81–95},
    numpages = {15}
}

@article{niu2023low,
  title={Low-loss interconnects for modular superconducting quantum processors},
  author={Niu, Jingjing and Zhang, Libo and Liu, Yang and Qiu, Jiawei and Huang, Wenhui and Huang, Jiaxiang and Jia, Hao and Liu, Jiawei and Tao, Ziyu and Wei, Weiwei and others},
  journal={Nature Electronics},
  volume={6},
  number={3},
  pages={235--241},
  year={2023},
  publisher={Nature Publishing Group UK London}
}

@inproceedings{liu_2023_tackling,
  author={Liu, Ji and Younis, Ed and Weiden, Mathias and Hovland, Paul and Kubiatowicz, John and Iancu, Costin},
  booktitle={2023 IEEE International Conference on Quantum Computing and Engineering (QCE)}, 
  title={Tackling the Qubit Mapping Problem with Permutation-Aware Synthesis}, 
  year={2023},
  volume={01},
  number={},
  pages={745-756},
  doi={10.1109/QCE57702.2023.00090}}

@article{gold2021entanglement,
  title={Entanglement across separate silicon dies in a modular superconducting qubit device},
  author={Gold, Alysson and Paquette, JP and Stockklauser, Anna and Reagor, Matthew J and Alam, M Sohaib and Bestwick, Andrew and Didier, Nicolas and Nersisyan, Ani and Oruc, Feyza and Razavi, Armin and others},
  journal={npj Quantum Information},
  volume={7},
  number={1},
  pages={142},
  year={2021},
  publisher={Nature Publishing Group UK London}
}

@article{bandic2024profiling,
  title={Profiling quantum circuits for their efficient execution on single-and multi-core architectures},
  author={Bandic, Medina and le Henaff, Pablo and Ovide, Anabel and Escofet, Pau and Ben Rached, Sahar and Rodrigo, Santiago and van Someren, Hans and Abadal, Sergi and Alarcon, Eduard and G. Almudever, Carmen and others},
  journal={Quantum Science and Technology},
  year = {2024}
}

@inproceedings{let_every_qubit,
author = {Lin, Sophia Fuhui and Sussman, Sara and Duckering, Casey and Mundada, Pranav S. and Baker, Jonathan M. and Kumar, Rohan S. and Houck, Andrew A. and Chong, Frederic T.},
title = {Let Each Quantum Bit Choose Its Basis Gates},
year = {2023},
isbn = {9781665462723},
publisher = {IEEE Press},
url = {https://doi.org/10.1109/MICRO56248.2022.00075},
doi = {10.1109/MICRO56248.2022.00075},
booktitle = {Proceedings of the 55th Annual IEEE/ACM International Symposium on Microarchitecture},
pages = {1042–1058},
numpages = {17},
location = {Chicago, Illinois, USA},
series = {MICRO '22}
}

@online{qiskit_basis_translation_errors,
	author = {},
	title = {{T}ranslation {E}rrors | {I}{B}{M} {Q}uantum {D}ocumentation --- docs.quantum.ibm.com},
	url = {https://docs.quantum.ibm.com/api/qiskit/qiskit.transpiler.passes.BasisTranslator#translation-errors},
	year = {},
	note = {[Accessed 2025-02-20]},
}

@book{mike_and_ike, 
place={Cambridge}, 
title={Quantum Computation and Quantum Information: 10th Anniversary Edition}, 
publisher={Cambridge University Press}, 
author={Nielsen, Michael A. and Chuang, Isaac L.}, 
year={2010},
doi={https://doi.org/10.1017/CBO9780511976667}
}

@article{vqe,
author={Peruzzo, Alberto
and McClean, Jarrod
and Shadbolt, Peter
and Yung, Man-Hong
and Zhou, Xiao-Qi
and Love, Peter J.
and Aspuru-Guzik, Al{\'a}n
and O'Brien, Jeremy L.},
title={A variational eigenvalue solver on a photonic quantum processor},
journal={Nature Communications},
year={2014},
month={Jul},
day={23},
volume={5},
number={1},
pages={4213},
issn={2041-1723},
doi={10.1038/ncomms5213},
url={https://doi.org/10.1038/ncomms5213}
}

@inproceedings{smith2019quantum,
  title={A quantum computational compiler and design tool for technology-specific targets},
  author={Smith, Kaitlin N and Thornton, Mitchell A},
  booktitle={Proceedings of the 46th International Symposium on Computer Architecture},
  pages={579--588},
  year={2019}
}

@INPROCEEDINGS{ovide2024sta,
  author={Ovide, Anabel and Cuomo, Daniele and Almudever, Carmen G.},
  booktitle={2024 IEEE International Conference on Quantum Computing and Engineering (QCE)}, 
  title={Scaling and Assigning Resources on ion Trap {QCCD} Architectures}, 
  year={2024},
  volume={01},
  number={},
  pages={959-970},
  doi={10.1109/QCE60285.2024.00115}}

@article{vanMeter_2010_DQC,
author = {Van Meter, Rodney and Ladd, Thaddeus D. and Fowler, Austin G. and Yamamoto, Yoshihisa},
title = {DISTRIBUTED QUANTUM COMPUTATION ARCHITECTURE USING SEMICONDUCTOR NANOPHOTONICS},
journal = {International Journal of Quantum Information},
volume = {08},
number = {01n02},
pages = {295-323},
year = {2010},
doi = {10.1142/S0219749910006435},
URL = {https://doi.org/10.1142/S0219749910006435},
eprint = {https://doi.org/10.1142/S0219749910006435}
}

@misc{gambetta_2024_IBMcouplers,
  title={{IBM} Quantum delivers on performance challenge made two years ago},
  author={Jay Gambetta and Ryan Mandelbaum},
  year={2024},
  howpublished={{IBM} Quantum Developer Conference (\url{https://www.ibm.com/quantum/blog/qdc-2024})},
}

@ARTICLE{Mandelbaum_2024_IBMCouplers,
  author={Mandelbaum, Ryan and Córcoles, Antonio D. and Gambetta, Jay},
  journal={IEEE Spectrum}, 
  title={{IBM}'s Big Bet on the Quantum-Centric Supercomputer: Recent Advances Point the Way to Useful Classical-Quantum Hybrids}, 
  year={2024},
  volume={61},
  number={9},
  pages={24-33},
  doi={10.1109/MSPEC.2024.10669253}
}

@article{LaRacuente2025modelingshortrange,
  doi = {10.22331/q-2025-01-08-1581},
  url = {https://doi.org/10.22331/q-2025-01-08-1581},
  title = {Modeling Short-Range Microwave Networks to Scale Superconducting Quantum Computation},
  author = {LaRacuente, Nicholas and Smith, Kaitlin N. and Imany, Poolad and Silverman, Kevin L. and Chong, Frederic T.},
  journal = {{Quantum}},
  issn = {2521-327X},
  publisher = {{Verein zur F{\"{o}}rderung des Open Access Publizierens in den Quantenwissenschaften}},
  volume = {9},
  pages = {1581},
  month = jan,
  year = {2025}
}

@misc{Nighthawk_topology,
  title={The superposition of impact and hype: building a quantum-powered future, one qubit at a time},
  author={{Economist Impact}},
  month={May},
  year={2025},
  howpublished={Available online at: \url{https://youtu.be/vvJvhoRPTgo?si=HGlxihFk_PON0ZfM&t=927}},
}

@article{Field_2024_RigettiModular,
  title = {Modular superconducting-qubit architecture with a multichip tunable coupler},
  author = {Field, Mark and Chen, Angela Q. and Scharmann, Ben and Sete, Eyob A. and Oruc, Feyza and Vu, Kim and Kosenko, Valentin and Mutus, Joshua Y. and Poletto, Stefano and Bestwick, Andrew},
  journal = {Phys. Rev. Appl.},
  volume = {21},
  issue = {5},
  pages = {054063},
  numpages = {9},
  year = {2024},
  month = {May},
  publisher = {American Physical Society},
  doi = {10.1103/PhysRevApplied.21.054063},
  url = {https://link.aps.org/doi/10.1103/PhysRevApplied.21.054063}
}

@misc{Rigetti_Cepheus,
  title={Rigetti makes its 36-qubit {C}epheus-1-36{Q} quantum computer available in the cloud},
  author={Trueman, Charlotte},
  month={August},
  year={2025},
  howpublished={Available online at: \url{https://www.datacenterdynamics.com/en/news/rigetti-makes-its-36-qubit-cepheus-1-36q-quantum-computer-available-in-the-cloud/}},
}

@misc{Rigetti_AspenM,
  title={Rigetti debuts multichip quantum processor with 80 qubits},
  author={{SiliconAngle}},
  month={June},
  year={2021},
  howpublished={Available online at: \url{https://siliconangle.com/2021/06/29/rigetti-looks-scale-quantum-computing-modular-processor-architecture/}},
}

@misc{Rigetti_2024_Roadmap,
  title={Investor Presentation},
  author={Rigetti},
  month={November},
  year={2024},
  howpublished={Available online at: \url{https://investors.rigetti.com/static-files/fbac3801-223f-4f0f-a207-47d25084a1d7}},
}

@misc{qiskit2024,
      title={Quantum computing with {Q}iskit},
      author={Javadi-Abhari, Ali and Treinish, Matthew and Krsulich, Kevin and Wood, Christopher J. and Lishman, Jake and Gacon, Julien and Martiel, Simon and Nation, Paul D. and Bishop, Lev S. and Cross, Andrew W. and Johnson, Blake R. and Gambetta, Jay M.},
      year={2024},
      doi={10.48550/arXiv.2405.08810},
      eprint={2405.08810},
      archivePrefix={arXiv},
      primaryClass={quant-ph}
}

@misc{IonQ_2024_Roadmap,
  title={{IonQ} Plots Path to Commercial (Quantum) Advantage},
  author={{HPCWire}},
  month={July},
  year={2024},
  howpublished={Available online at: \url{https://www.hpcwire.com/2024/07/02/ionq-plots-path-to-commercial-quantum-advantage/}},
}

@misc{IBM_2025_FTQC,
  title={How {IBM} will build the world's first large-scale, fault-tolerant quantum computer},
  author={Ryan Mandelbaum and 
          Jay Gambetta and 
          Jerry Chow and 
          Tushar Mittal and
          Theodore J. Yoder and 
          Andrew Cross and
          Matthias Steffen},
  month={Jun},
  year={2025},
  howpublished={Available online at: \url{https://www.ibm.com/quantum/blog/large-scale-ftqc/}},
}

@inproceedings{Tannu_2019_ErrorVariability,
author = {Tannu, Swamit S. and Qureshi, Moinuddin K.},
title = {Not All Qubits Are Created Equal: A Case for Variability-Aware Policies for {NISQ}-Era Quantum Computers},
year = {2019},
isbn = {9781450362405},
publisher = {Association for Computing Machinery},
address = {New York, NY, USA},
url = {https://doi.org/10.1145/3297858.3304007},
doi = {10.1145/3297858.3304007},
booktitle = {Proceedings of the Twenty-Fourth International Conference on Architectural Support for Programming Languages and Operating Systems},
pages = {987–999},
numpages = {13},
location = {Providence, RI, USA},
series = {ASPLOS '19}
}

@misc{DARPA_HARQ,
  title={Heterogeneous Architectures for Quantum ({HARQ})},
  author={{DARPA}},
  month={August},
  year={2025},
  howpublished={Available online at: \url{https://sam.gov/opp/967cd8f4c3554b448d7ba3325ed99de2/view}},
}

@inproceedings{Murali_2019_NoiseMapping,
author = {Murali, Prakash and Baker, Jonathan M. and Javadi-Abhari, Ali and Chong, Frederic T. and Martonosi, Margaret},
title = {Noise-Adaptive Compiler Mappings for Noisy Intermediate-Scale Quantum Computers},
year = {2019},
isbn = {9781450362405},
publisher = {Association for Computing Machinery},
address = {New York, NY, USA},
url = {https://doi.org/10.1145/3297858.3304075},
doi = {10.1145/3297858.3304075},
booktitle = {Proceedings of the Twenty-Fourth International Conference on Architectural Support for Programming Languages and Operating Systems},
pages = {1015–1029},
numpages = {15},
location = {Providence, RI, USA},
series = {ASPLOS '19}
}

@article{dasgupta2021stability,
  title={Stability of noisy quantum computing devices},
  author={Dasgupta, Samudra and Humble, Travis S},
  journal={arXiv preprint arXiv:2105.09472},
  year={2021}
}

@INPROCEEDINGS{10228915,
  author={Mao, Yingling and Liu, Yu and Yang, Yuanyuan},
  booktitle={IEEE INFOCOM 2023 - IEEE Conference on Computer Communications}, 
  title={Qubit Allocation for Distributed Quantum Computing}, 
  year={2023},
  volume={},
  number={},
  pages={1-10},
  doi={10.1109/INFOCOM53939.2023.10228915}}

@article{10.1145/3730585,
author = {Liu, Lei},
title = {Lar{Q}ucut: A New Cutting and Mapping Approach for Large-sized Quantum Circuits in Distributed Quantum Computing (DQC) Environments},
year = {2025},
publisher = {Association for Computing Machinery},
address = {New York, NY, USA},
issn = {1544-3566},
url = {https://doi.org/10.1145/3730585},
doi = {10.1145/3730585},
note = {Just Accepted},
journal = {ACM Trans. Archit. Code Optim.},
month = apr
}

@INPROCEEDINGS{10821229,
  author={Burt, Felix and Chen, Kuan-Cheng and Leung, Kin K.},
  booktitle={2024 IEEE International Conference on Quantum Computing and Engineering (QCE)}, 
  title={Generalised Circuit Partitioning for Distributed Quantum Computing}, 
  year={2024},
  volume={02},
  number={},
  pages={173-178},
  doi={10.1109/QCE60285.2024.10273}}

@INPROCEEDINGS {10313645,
author = { Davis, Marc G. and Chung, Joaquin and Englund, Dirk and Kettimuthu, Rajkumar },
booktitle = { 2023 IEEE International Conference on Quantum Computing and Engineering (QCE) },
title = {Towards Distributed Quantum Computing by Qubit and Gate Graph Partitioning Techniques},
year = {2023},
volume = {},
ISSN = {},
pages = {161-167},
doi = {10.1109/QCE57702.2023.00026},
url = {https://doi.ieeecomputersociety.org/10.1109/QCE57702.2023.00026},
publisher = {IEEE Computer Society},
address = {Los Alamitos, CA, USA},
month =sep}

@article{10.1145/3579367,
author = {Cuomo, Daniele and Caleffi, Marcello and Krsulich, Kevin and Tramonto, Filippo and Agliardi, Gabriele and Prati, Enrico and Cacciapuoti, Angela Sara},
title = {Optimized Compiler for Distributed Quantum Computing},
year = {2023},
issue_date = {June 2023},
publisher = {Association for Computing Machinery},
address = {New York, NY, USA},
volume = {4},
number = {2},
url = {https://doi.org/10.1145/3579367},
doi = {10.1145/3579367},
journal = {ACM Transactions on Quantum Computing},
month = feb,
articleno = {15},
numpages = {29}
}

@article{Kreppel2023quantumcircuit,
  doi = {10.22331/q-2023-11-08-1176},
  url = {https://doi.org/10.22331/q-2023-11-08-1176},
  title = {Quantum Circuit Compiler for a Shuttling-Based Trapped-Ion Quantum Computer},
  author = {Kreppel, Fabian and Melzer, Christian and Olvera Mill{\'{a}}n, Diego and Wagner, Janis and Hilder, Janine and Poschinger, Ulrich and Schmidt-Kaler, Ferdinand and Brinkmann, Andr{\'{e}}},
  journal = {{Quantum}},
  issn = {2521-327X},
  publisher = {{Verein zur F{\"{o}}rderung des Open Access Publizierens in den Quantenwissenschaften}},
  volume = {7},
  pages = {1176},
  month = nov,
  year = {2023}
}

@inproceedings{10.1145/3526241.3530366,
author = {Upadhyay, Suryansh and Saki, Abdullah Ash and Topaloglu, Rasit Onur and Ghosh, Swaroop},
title = {A Shuttle-Efficient Qubit Mapper for Trapped-Ion Quantum Computers},
year = {2022},
isbn = {9781450393225},
publisher = {Association for Computing Machinery},
address = {New York, NY, USA},
url = {https://doi.org/10.1145/3526241.3530366},
doi = {10.1145/3526241.3530366},
booktitle = {Proceedings of the Great Lakes Symposium on VLSI 2022},
pages = {305–308},
numpages = {4},
location = {Irvine, CA, USA},
series = {GLSVLSI '22}
}

@inproceedings{hetarch,
author = {Stein, Samuel and Sussman, Sara and Tomesh, Teague and Guinn, Charles and Tureci, Esin and Lin, Sophia Fuhui and Tang, Wei and Ang, James and Chakram, Srivatsan and Li, Ang and Martonosi, Margaret and Chong, Fred and Houck, Andrew A. and Chuang, Isaac L. and Demarco, Michael},
title = {{H}et{A}rch: Heterogeneous Microarchitectures for Superconducting Quantum Systems},
year = {2023},
isbn = {9798400703294},
publisher = {Association for Computing Machinery},
address = {New York, NY, USA},
url = {https://doi.org/10.1145/3613424.3614300},
doi = {10.1145/3613424.3614300},
booktitle = {Proceedings of the 56th Annual IEEE/ACM International Symposium on Microarchitecture},
pages = {539–554},
numpages = {16},
location = {Toronto, ON, Canada},
series = {MICRO '23}
}

@article{Xiang_superconducting_hybrid_2013,
  title = {Hybrid quantum circuits: Superconducting circuits interacting with other quantum systems},
  author = {Xiang, Ze-Liang and Ashhab, Sahel and You, J. Q. and Nori, Franco},
  journal = {Rev. Mod. Phys.},
  volume = {85},
  issue = {2},
  pages = {623--653},
  numpages = {0},
  year = {2013},
  month = {Apr},
  publisher = {American Physical Society},
  doi = {10.1103/RevModPhys.85.623},
  url = {https://link.aps.org/doi/10.1103/RevModPhys.85.623}
}

@article{
Kurizki_hybrid_2015,
author = {Gershon Kurizki  and Patrice Bertet  and Yuimaru Kubo  and Klaus Mølmer  and David Petrosyan  and Peter Rabl  and Jörg Schmiedmayer },
title = {Quantum technologies with hybrid systems},
journal = {Proceedings of the National Academy of Sciences},
volume = {112},
number = {13},
pages = {3866-3873},
year = {2015},
doi = {10.1073/pnas.1419326112},
URL = {https://www.pnas.org/doi/abs/10.1073/pnas.1419326112},
eprint = {https://www.pnas.org/doi/pdf/10.1073/pnas.1419326112}}

@article{PhysRevLett.114.080501,
  title = {High-Contrast Qubit Interactions Using Multimode Cavity {QED}},
  author = {McKay, David C. and Naik, Ravi and Reinhold, Philip and Bishop, Lev S. and Schuster, David I.},
  journal = {Phys. Rev. Lett.},
  volume = {114},
  issue = {8},
  pages = {080501},
  numpages = {5},
  year = {2015},
  month = {Feb},
  publisher = {American Physical Society},
  doi = {10.1103/PhysRevLett.114.080501},
  url = {https://link.aps.org/doi/10.1103/PhysRevLett.114.080501}
}

@article{Cia_2024,
author = {Chia, Cleaven and Huang, Ding and Leong, Victor and Kong, Jian Feng and Goh, Kuan Eng Johnson},
title = {Hybrid Quantum Systems with Artificial Atoms in Solid State},
journal = {Advanced Quantum Technologies},
volume = {7},
number = {5},
pages = {2300461},
doi = {https://doi.org/10.1002/qute.202300461},
url = {https://advanced.onlinelibrary.wiley.com/doi/abs/10.1002/qute.202300461},
eprint = {https://advanced.onlinelibrary.wiley.com/doi/pdf/10.1002/qute.202300461},
year = {2024}
}

@Article{Zhou2014,
author={Zhou, Jian
and Hu, Yong
and Yin, Zhang-qi
and Wang, Z. D.
and Zhu, Shi-Liang
and Xue, Zheng-Yuan},
title={High fidelity quantum state transfer in electromechanical systems with intermediate coupling},
journal={Scientific Reports},
year={2014},
month={Aug},
day={29},
volume={4},
number={1},
pages={6237},
issn={2045-2322},
doi={10.1038/srep06237},
url={https://doi.org/10.1038/srep06237}
}

@article{PhysRevLett.108.130504,
  title = {Quantum Interface between an Electrical Circuit and a Single Atom},
  author = {Kielpinski, D. and Kafri, D. and Woolley, M. J. and Milburn, G. J. and Taylor, J. M.},
  journal = {Phys. Rev. Lett.},
  volume = {108},
  issue = {13},
  pages = {130504},
  numpages = {4},
  year = {2012},
  month = {Mar},
  publisher = {American Physical Society},
  doi = {10.1103/PhysRevLett.108.130504},
  url = {https://link.aps.org/doi/10.1103/PhysRevLett.108.130504}
}

@Article{Clerk2020,
author={Clerk, A. A.
and Lehnert, K. W.
and Bertet, P.
and Petta, J. R.
and Nakamura, Y.},
title={Hybrid quantum systems with circuit quantum electrodynamics},
journal={Nature Physics},
year={2020},
month={Mar},
day={01},
volume={16},
number={3},
pages={257-267},
issn={1745-2481},
doi={10.1038/s41567-020-0797-9},
url={https://doi.org/10.1038/s41567-020-0797-9}
}

@Article{Yu2016,
author={Yu, Deshui
and Valado, Mar{\'i}a Mart{\'i}nez
and Hufnagel, Christoph
and Kwek, Leong Chuan
and Amico, Luigi
and Dumke, Rainer},
title={Quantum State Transmission in a Superconducting Charge Qubit-Atom Hybrid},
journal={Scientific Reports},
year={2016},
month={Dec},
day={06},
volume={6},
number={1},
pages={38356},
issn={2045-2322},
doi={10.1038/srep38356},
url={https://doi.org/10.1038/srep38356}
}

@article{Daniilidis_2013,
doi = {10.1088/1367-2630/15/7/073017},
url = {https://doi.org/10.1088/1367-2630/15/7/073017},
year = {2013},
month = {jul},
publisher = {IOP Publishing},
volume = {15},
number = {7},
pages = {073017},
author = {Daniilidis, Nikos and Gorman, Dylan J and Tian, Lin and Häffner, Hartmut},
title = {Quantum information processing with trapped electrons and superconducting electronics},
journal = {New Journal of Physics}
}

@misc{HQAN,
  title={Hybrid Quantum Architectures and Networks ({HQAN})},
  author={{University of Illinois Urbana-Champaign}},
  month={September},
  year={2020},
  howpublished={Available online at: \url{https://hqan.illinois.edu/}},
}

@misc{piccinelli_chemistry,
      title={Quantum chemistry with provable convergence via randomized sample-based quantum diagonalization}, 
      author={Samuele Piccinelli and Alberto Baiardi and Max Rossmannek and Almudena Carrera Vazquez and Francesco Tacchino and Stefano Mensa and Edoardo Altamura and Ali Alavi and Mario Motta and Javier Robledo-Moreno and William Kirby and Kunal Sharma and Antonio Mezzacapo and Ivano Tavernelli},
      year={2025},
      eprint={2508.02578},
      archivePrefix={arXiv},
      primaryClass={quant-ph},
      url={https://arxiv.org/abs/2508.02578}, 
}

@article{Kim2023preFTUtility,
  title        = {Evidence for the utility of quantum computing before fault tolerance},
  author       = {Youngseok Kim and Andrew Eddins and Sajant Anand and Ken Xuan Wei
                  and Ewout van den Berg and Sami Rosenblatt and Hasan Nayfeh
                  and Yantao Wu and Michael Zaletel and Kristan Temme and Abhinav Kandala},
  journal      = {Nature},
  volume       = {618},
  number       = {7965},
  pages        = {500--505},
  year         = {2023},
  doi          = {10.1038/s41586-023-06096-3},
}

@article{sciadv_utility_2025,
    author = {Javier Robledo-Moreno  and Mario Motta  and Holger Haas  and Ali Javadi-Abhari  and Petar Jurcevic  and William Kirby  and Simon Martiel  and Kunal Sharma  and Sandeep Sharma  and Tomonori Shirakawa  and Iskandar Sitdikov  and Rong-Yang Sun  and Kevin J. Sung  and Maika Takita  and Minh C. Tran  and Seiji Yunoki  and Antonio Mezzacapo },
    title = {Chemistry beyond the scale of exact diagonalization on a quantum-centric supercomputer},
    journal = {Science Advances},
    volume = {11},
    number = {25},
    pages = {eadu9991},
    year = {2025},
    doi = {10.1126/sciadv.adu9991},
    URL = {https://www.science.org/doi/abs/10.1126/sciadv.adu9991},
}

@article{Abanin2025ConstructiveInterference,
  title        = {Observation of constructive interference at the edge of quantum ergodicity},
  author       = {Abanin, Dmitry A. and Acharya, Rajeev and Aghababaie-Beni, Laleh and Aigeldinger, Georg and Ajoy, Ashok and Alcaraz, Ross and Aleiner, Igor and Andersen, Trond I. and Ansmann, Markus and Arute, Frank and Arya, Kunal and Asfaw, Abraham and Astrakhantsev, Nikita and Atalaya, Juan and Babbush, Ryan and Bacon, Dave and Ballard, Brian and Bardin, Joseph C. and Bengs, Christian and Bengtsson, Andreas and Bilmes, Alexander and Boixo, Sergio and Bortoli, Gina and Bourassa, Alexandre and Bovaird, Jenna and Bowers, Dylan and Brill, Leon and Broughton, Michael and Browne, David and Buchea, Brett and Buckley, Bob and Buell, David A. and Burger, Tim and Burkett, Brian and Bushnell, Nicholas and Cabrera, Anthony and Campero, Juan and Chang, Hung-Shen and Chen, Yu and Chen, Zijun and Chiaro, Ben and Chih, Liang-Ying and Chik, Desmond and Chou, Charina and Claes, Jahan and Cleland, Agnetta Y. and Cogan, Josh and Cohen, Saul and Collins, Roberto and Conner, Paul and Courtney, William and Crook, Alexander L. and Curtin, Ben and Das, Sayan and De Lorenzo, Laura and Debroy, Dripto M. and Demura, Sean and Devoret, Michel and Di Paolo, Agustin and Donohoe, Paul and Drozdov, Ilya and Dunsworth, Andrew and Earle, Clint and Eickbusch, Alec and Elbag, Aviv Moshe and Elzouka, Mahmoud and Erickson, Catherine and Faoro, Lara and Farhi, Edward and Ferreira, Vinicius S. and Flores Burgos, Leslie and Forati, Ebrahim and Fowler, Austin G. and Foxen, Brooks and Ganjam, Suhas and Garcia, Gonzalo and Gasca, Robert and Genois, Elie and Giang, William and Gidney, Craig and Gilboa, Dar and Gosula, Raja and Grajales Dau, Alejandro and Graumann, Dietrich and Greene, Alex and Gross, Jonathan A. and Gu, Hanfeng and Habegger, Steve and Hall, John and Hamamura, Ikko and Hamilton, Michael C. and Hansen, Monica and Harrigan, Matthew P. and Harrington, Sean D. and Heslin, Stephen and Heu, Paula and Higgott, Oscar and Hill, Gordon and Hilton, Jeremy and Hong, Sabrina and Huang, Hsin-Yuan and Huff, Ashley and Huggins, William J. and Ioffe, Lev B. and Isakov, Sergei V. and Iveland, Justin and Jeffrey, Evan and Jiang, Zhang and Jin, Xiaoxuan and Jones, Cody and Jordan, Stephen and Joshi, Chaitali and Juhas, Pavol and Kabel, Andreas and Kafri, Dvir and Kang, Hui and Karamlou, Amir H. and Kechedzhi, Kostyantyn and Kelly, Julian and Khaire, Trupti and Khattar, Tanuj and Khezri, Mostafa and Kim, Seon and King, Robbie and Klimov, Paul V. and Klots, Andrey R. and Kobrin, Bryce and Korotkov, Alexander N. and Kostritsa, Fedor and Kothari, Robin and Kreikebaum, John Mark and Kurilovich, Vladislav D. and Kyoseva, Elica and Landhuis, David and Lange-Dei, Tiano and Langley, Brandon W. and Laptev, Pavel and Lau, Kim-Ming and Le Guevel, Loick and Ledford, Justin and Lee, Joonho and Lee, Kenny and Lensky, Yuri D. and Leon, Shannon and Lester, Brian J. and Li, Wing Yan and Lill, Alexander T. and Liu, Wayne and Livingston, William P. and Locharla, Aditya and Lucero, Erik and Lundahl, Daniel and Lunt, Aaron and Madhuk, Sid and Malone, Fionn D. and Maloney, Ashley and Mandra, Salvatore and Manyika, James M. and Martin, Leigh S. and Martin, Orion and Martin, Steven and Matias, Yossi and Maxfield, Cameron and McClean, Jarrod R. and McEwen, Matt and Meeks, Seneca and Megrant, Anthony and Mi, Xiao and Miao, Kevin C. and Mieszala, Amanda and Molavi, Reza and Molina, Sebastian and Montazeri, Shirin and Morvan, Alexis and Movassagh, Ramis and Mruczkiewicz, Wojciech and Naaman, Ofer and Neeley, Matthew and Nersisyan, Ani and Neven, Hartmut and Newman, Michael and Ng, Jiun How and Nguyen, Anthony and Nguyen, Murray and Ni, Chia-Hung and Niu, Murphy Yuezhen and Oas, Logan and O'Brien, Thomas E. and Oliver, William D. and Opremcak, Alex and Ottosson, Kristoffer and Petukhov, Andre and Pizzuto, Alex and Platt, John and Potter, Rebecca and Pritchard, Orion and Pryadko, Leonid P. and Quintana, Chris and Ramachandran, Ganesh and Ramanathan, Chandrasekhar and Reagor, Matthew J. and Redding, John and Rhodes, David M. and Roberts, Gabrielle and Rosenberg, Eliott and Rosenfeld, Emma and Roushan, Pedram and Rubin, Nicholas C. and Saei, Negar and Sank, Daniel and Sankaragomathi, Kannan and Satzinger, Kevin J. and Schmidhuber, Alexander and Schurkus, Henry F. and Schuster, Christopher and Shearn, Michael J. and Shorter, Aaron and Shutty, Noah and Shvarts, Vladimir and Sivak, Volodymyr and Skruzny, Jindra and Small, Spencer and Smelyanskiy, Vadim and Smith, W. Clarke and Somma, Rolando D. and Springer, Sofia and Sterling, George and Strain, Doug and Suchard, Jordan and Suchsland, Philippe and Szasz, Aaron and Sztein, Alex and Thor, Douglas and Tomita, Eifu and Torres, Alfredo and Torunbalci, M. Mert and Vaishnav, Abeer and Vargas, Justin and Vdovichev, Sergey and Vidal, Guifre and Villalonga, Benjamin and Vollgraff Heidweiller, Catherine and Waltman, Steven and Wang, Shannon X. and Ware, Brayden and Weber, Kate and Weidel, Travis and Westerhout, Tom and White, Theodore and Wong, Kristi and Woo, Bryan W. K. and Xing, Cheng and Yao, Z. Jamie and Yeh, Ping and Ying, Bicheng and Yoo, Juhwan and Yosri, Noureldin and Young, Grayson and Zalcman, Adam and Zhang, Chongwei and Zhang, Yaxing and Zhu, Ningfeng and Zobrist, Nicholas},
  journal      = {Nature},
  volume       = {646},
  number       = {8086},
  pages        = {825--830},
  year         = {2025},
  doi          = {10.1038/s41586-025-09526-6}
}
